\documentclass[%
 reprint,
 amsmath,amssymb,
 aps,
 prb,
]{revtex4-1}

\usepackage{graphicx}
\usepackage{dcolumn}
\usepackage{bm}
\usepackage{braket}
\usepackage{color}

\begin{document}

\title{Drastic enhancement of the thermal Hall angle in a $d$-wave superconductor}

\author{Hikaru Ueki,$^1$
Hiroki Morita,$^2$
Marie Ohuchi,$^2$ 
and Takafumi Kita$^2$}
\affiliation{%
$^1$Department of Mathematics and Physics, Hirosaki University, Hirosaki, Aomori 036-8561, Japan \\
$^2$Department of Physics, Hokkaido University, Sapporo, Hokkaido 060-0810, Japan
}%

\date{\today}

\begin{abstract} 
A drastic enhancement of the thermal Hall angle in $d$-wave superconductors 
was observed experimentally in a cuprate superconductor and in CeCoIn$_5$  
at low temperatures and very weak magnetic field 
[Phys. Rev. Lett. {\bf 86}, 890 (2001); Phys. Rev. B {\bf 72}, 214515 (2005)]. 
However, to the best of our knowledge, its microscopic calculation has not been performed yet.
To study this microscopically, 
we derive the thermal Hall coefficient in extreme type-II superconductors 
with an isolated pinned vortex 
based on the augmented quasiclassical equations of superconductivity with the Lorentz force. 
Using it, we can confirm that  
the quasiparticle relaxation time and the thermal Hall angle are enhanced 
in $d$-wave superconductors 
without impurities of the resonant scattering 
because quasiparticles around the gap nodes which become dominant near zero temperature 
are restricted to the momentum in a specific orientation. 
This enhancement of the thermal Hall angle may also be observed in other nodal superconductors 
with large magnetic-penetration depth.

\end{abstract}

\pacs{Valid PACS appear here}

\maketitle


\section{Introduction \label{sec:I}}
It was observed experimentally 
that the thermal conductivity and the thermal Hall angle are greatly enhanced 
in the superconducting states of 
cuprates \cite{Cohn,Krishana95,Zeini99,Krishana99,Zhang00,Ocana,Zhang01,Zeini01} 
and CeCoIn$_5$ \cite{Kasahara}.
These materials are regarded as $d$-wave superconductors, 
and their thermal Hall conductivity is also consistent with 
the scaling relation for a $d$-wave pairing 
based on the Bogoliubov--de Gennes equation \cite{Simon}. 
Hirschfeld {\it et al.} also calculated the longitudinal thermal conductivity 
in a $d$-wave superconductor, 
including the effect of antiferromagnetic spin fluctuations 
within the random phase approximation (RPA), 
and studied the large peak structure of the longitudinal thermal conductivity 
in YBa$_2$Cu$_3$O$_{7-\delta}$ (YBCO) 
\cite{Hirschfeld96}.
The enhancement of the thermal Hall angle 
has been explained to originate from the enhancement of the quasiparticle mean free path 
\cite{Zhang01,Kasahara}.
However, to the best of our knowledge, its microscopic calculation has not been carried out yet. 
Therefore, the origin of the enhancement of the quasiparticle mean free path needs to be clarified 
using a fully microscopic treatment. 
The purpose of this present paper is to develop a theoretical formalism for investigating 
the thermal Hall conductivity microscopically within the quasiclassical theory
and to study this drastic enhancement of the thermal Hall angle 
in the superconducting states of YBCO and CeCoIn$_5$.

Numerous theoretical studies have been carried out, 
such as on the longitudinal component of the thermal conductivities 
in both conventional ($s$-wave) \cite{Bardeen,Graf,Adachi} 
and unconventional superconductors 
\cite{Hirschfeld96,Schmitt,Hirschfeld88,Graf,Kubert,Franz,Vekhter,Vorontsov,Adachi,Hara,Kobayashi}
and the spontaneous thermal Hall conductivity due to 
the intrinsic \cite{Read,Nomura,Sumiyoshi,Imai16,Yoshioka}
and extrinsic \cite{Yip,Yilmaz,Ngampruetikorn,Ngampruetikorn20} 
mechanisms in chiral superconductors. 
On the other hand, 
several studies have been carried out on the thermal Hall conductivity 
in $d$-wave superconductors, 
in terms of the cross section of quasiparticle scattering from a single vortex \cite{Durst}, 
due to the Berry phase acquired by a quasiparticle in the vortex lattice state 
\cite{Vafek01,Cvetkovic,Murray,Vafek15}, 
based on the Kubo formula approach and taking into account the Doppler shift effect 
\cite{Shahzamanian}. 
The thermal Hall conductivity due to the Lorentz force in type-II superconductors 
is not fully understood, this may be because the Lorentz force is missing 
from the quasiclassical Eilenberger equations \cite{Eilenberger}, 
which is a powerful tool for investigating inhomogeneous and nonequilibrium superconductors microscopically \cite{Eschrig}. 
More precisely, the component of the magnetic Lorentz force balanced with the Hall electric field 
and/or force induced by a transverse temperature gradient 
may be missing from the standard Eilenberger equations, 
since there is the component of the magnetic Lorentz force balanced 
with the hydrodynamic force in the Ginzburg--Landau equations \cite{Kato16}.

Here we derive the thermal Hall coefficient in extreme type-II superconductors 
with an isolated pinned vortex,
based on the augmented quasiclassical equations of superconductivity 
with the Lorentz force in the Keldysh formalism \cite{Kita01} 
by incorporating a next-to-leading-order contribution 
in the expansion of the Gor'kov equations \cite{Gor'kov,Gor'kov2}
in terms of the quasiclassical parameter $\delta \equiv 1 / k_{\rm F} \xi_0(0)$, 
where $k_{\rm F}$ and $\xi_0(0)$ denote the Fermi wave number 
and the coherence length at zero temperature, respectively. 
The calculations of linear responses based on the augmented equations 
obtained from this derivation \cite{Kita01}
have not been performed yet, as far as we know, except for the flux-flow Hall effect \cite{Arahata}.
In extreme type-II superconductors with an isolated pinned vortex, 
the thermal conductivity is dominated by the contribution from quasiparticles outside the core. 
Thus, we consider the contribution of the Doppler shifted quasiparticles 
due to the circulating supercurrent around the core (i.e., the Volovik effect) \cite{Kubert,Volovik,Kopnin}  
and neglect that of the Andreev-reflected quasiparticles in the core \cite{Dahm,Nagai}. 
The newly derived expression for the thermal conductivity is an extension of the expression 
proposed by Graf {\it et al.} \cite{Graf} 
and can also describe the thermal Hall effect in both fully gapped and nodal superconductors. 
Based on this formalism, 
we calculate the thermal conductivity and the thermal Hall angle 
in $d$-wave superconductors 
and address the enhancement of $\lim_{H \to 0} \tan \theta_{\rm H} / H$ 
at low temperatures in YBCO and CeCoIn$_5$ measured 
by Zhang {\it et al.} in Ref. \onlinecite{Zhang01} and by Kasahara {\it et al.} 
in Ref. \onlinecite{Kasahara}, 
where $\tan \theta_{\rm H}$ and $H$ denote the thermal Hall angle and the external magnetic field, 
respectively. 
It is also imperative to consider impurity scattering close to the unitarity limit 
to explain the experimental results on
the longitudinal thermal conductivity in heavy-fermion superconductors and Zn-doped YBCO
\cite{Coffey,Hirschfeld96,Schmitt,Hirschfeld88}. 
Thus, we also study the impurity effect on the thermal Hall conductivity and the thermal Hall angle
in a $d$-wave superconductor, 
using the impurity self-energy by $t$-matrix approximation \cite{Hirschfeld96,Schmitt,Hirschfeld88,Graf,Vorontsov,Kato00}.

This paper is organized as follows. 
In Sec. \ref{sec:II}, we present the augmented quasiclassical equations of superconductivity 
with the Lorentz force in the Keldysh formalism. 
In Sec. \ref{sec:III}, we derive the thermal Hall coefficient 
in extreme type-II superconductors with an isolated pinned vortex
based on the augmented quasiclassical equations of superconductivity with the Lorentz force and the linear response theory. 
In Sec. \ref{sec:IV}, we present numerical results for the thermal Hall effect in $d$-wave superconductors. 
In Sec. \ref{sec:V}, we provide concluding remarks.

\section{Augmented Quasiclassical Equations \label{sec:II}} 
We consider type-II superconductors with pinned vortices 
and neglect the pair-potential-gradient force \cite{Ohuchi,Masaki17,Masaki18} 
and the pressure difference arising from the slope in the density of states (DOS) \cite{Ueki18}, 
which only contribute to the vortex-core charging in this case, 
since the charge in a pinned vortex does not contribute to thermal conductivity.  
For simplicity, we also restrict ourselves to the spin-singlet pairing without spin paramagnetism. 
Then the quasiclassical equations of superconductivity with the Lorentz force are given in the Keldysh formalism by \cite{Kita01}
\begin{align}
&\left[ \varepsilon \check{\tau}_3 - \check{\Delta} - \check{\sigma}, \check{g} \right]_\circ 
+ {i} \hbar {\bm v}_{\rm F} \cdot {\bm \partial} \check{g} \notag \\
&+ \frac{{i}\hbar}{2} \left[ e {\bm v}_{\rm F} \cdot {\bm E} \frac{\partial}{\partial \varepsilon}
+ e \left( {\bm v}_{\rm F} \times {\bm B} \right) \cdot \frac{\partial}{\partial {\bm p}_{\rm F}} \right] 
\left\{ \check{\tau}_3, \check{g} \right\} = \check{0}, \label{AQCEq}
\end{align}
where $e < 0$ is the electron charge, 
${\bm E}$ is the electric field, 
${\bm B}$ is the magnetic field, 
$\varepsilon$ is the excitation energy, 
${\bm v}_{\rm F}$ is the Fermi velocity,
and ${\bm p}_{\rm F}$ is the Fermi momentum.
The Green's functions $\check{g}$, the pair potential $\check{\Delta}$, 
and the Born-type impurity self-energy $\check{\sigma}$ can be written as
\begin{align}
&\check{g} =
\begin{bmatrix}
\hat{g}^{\rm R} & \hat{g}^{\rm K} \\
\hat{0} & \hat{g}^{\rm A}
\end{bmatrix}, \ \ \ 
\check{\Delta} =
\begin{bmatrix}
\hat{\Delta} & \hat{0} \\
\hat{0} & \hat{\Delta}
\end{bmatrix}, \notag \\  
&\check{\sigma} = 
- \frac{{i} \hbar}{2 \tau} \langle \check{g} \rangle_{\rm F} = 
\begin{bmatrix}
\hat{\sigma}^{\rm R} & \hat{\sigma}^{\rm K} \\
\hat{0} & \hat{\sigma}^{\rm A}
\end{bmatrix}, 
\end{align}
where $\tau$ is the relaxation time 
and $\langle \cdots \rangle_{\rm F}$ denotes the Fermi surface average 
with $\langle 1 \rangle_{\rm F} = 1$. 
This Born-type impurity self-energy will later be rewritten to the impurity self-energy by the $t$-matrix approximation \cite{Graf,Vorontsov,Kato00}.
The $2 \times 2$ retarded and Keldysh Green's functions 
and the $2 \times 2$ pair potential can be written as
\begin{align}
&\hat{g}^{\rm R,A} =
\begin{bmatrix}
g^{\rm R,A} & - {i} f^{\rm R,A} \\
{i} \bar{f}^{\rm R,A} & - \bar{g}^{\rm R,A}
\end{bmatrix}, \ \ \ 
\hat{g}^{\rm K} =
\begin{bmatrix}
g^{\rm K} & - {i} f^{\rm K} \\
- {i} \bar{f}^{\rm K} & \bar{g}^{\rm K}
\end{bmatrix}, \notag \\ 
&\hat{\Delta} =
\begin{bmatrix}
0 & - \Delta \phi \\
\Delta^* \phi^* & 0
\end{bmatrix}, 
\end{align}
where the barred function in the Keldysh formalism is defined generally by
$
\bar{X} (\varepsilon, {\bm p}_{\rm F}, {\bm r}, t) 
\equiv X^* (- \varepsilon, - {\bm p}_{\rm F}, {\bm r}, t), 
$
$\Delta = \Delta ({\bm r}, t)$ denotes the amplitude of the energy gap, 
and $\phi = \phi ({\bm p}_{\rm F})$ is the basis function on the Fermi surface 
normalized as $\langle | \phi |^2 \rangle_{\rm F} = 1$.
Matrix $\check{\tau}_3$ is given by 
\begin{align}
\check{\tau}_3 =
\begin{bmatrix}
\hat{\tau}_3 & \hat{0} \\
\hat{0} & \hat{\tau}_3
\end{bmatrix}, \ \ \ 
\hat{\tau}_3 =
\begin{bmatrix}
1 & 0 \\
0 & - 1
\end{bmatrix}. 
\end{align}
The Green's functions satisfy the following symmetry relations: 
\begin{align}
\hat{g}^{\rm A} = - \hat{\tau}_3 \hat{g}^{{\rm R}\dagger} \hat{\tau}_3, \ \ \ 
\hat{g}^{\rm K} = \hat{\tau}_3 \hat{g}^{{\rm K}\dagger} \hat{\tau}_3. \label{symG}
\end{align}

We choose the gauge 
${\bm E} = - \partial {\bm A} / \partial t$ and ${\bm B} = {\bm \nabla} \times {\bm A}$ with $\Phi = 0$, 
where ${\bm A}$ and $\Phi$ denote the vector and scalar potentials. 
Notations $[a, b]_\circ$ and $\{a, b \}$ are then given by 
$[a, b]_\circ \equiv a \circ b - b \circ a$ and $\{a, b \} \equiv a b + b a$ with 
\begin{align}
&a \circ b \notag \\
&\equiv \exp \left[ \frac{ i \hbar}{2} 
\left( \frac{\partial}{\partial \varepsilon} \frac{\partial}{\partial t'} 
- \frac{\partial}{\partial t} \frac{\partial}{\partial \varepsilon'} \right)  \right] 
a (\varepsilon, t) b (\varepsilon', t')\bigg|_
{\varepsilon' = \varepsilon, t' = t}.
\end{align}
The gauge invariant derivative ${\bm \partial}$ is defined by 
\begin{align}
{\bm \partial} \equiv \left\{ \begin{array}{ll} {\bm \nabla} &  {\rm on} \  g^{\rm R,A,K}, \bar{g}^{\rm R,A,K} \\
\displaystyle {\bm \nabla} - i \frac{2 e {\bm A}}{\hbar} &  {\rm on} \ f^{\rm R,A,K}, \Delta \\
\displaystyle  {\bm \nabla} + i \frac{2 e {\bm A}}{\hbar} & {\rm on} \ \bar{f}^{\rm R,A,K}, \bar{\Delta}
\end{array}\right. .
\end{align}

The equation of the gap amplitude for the weak-coupling limit is given by 
\begin{align}
\Delta = 
\frac{{\rm g}_0}{4 i} \int_{- \varepsilon_{\rm c}}^{\varepsilon_{\rm c}} 
\langle f^{\rm K} \phi^* \rangle_{\rm F} d \varepsilon, 
\label{gapeq}
\end{align}
where $\varepsilon_{\rm c}$ and ${\rm g}_0$ 
denote the cutoff energy and the coupling constant, respectively, defined by
${\rm g}_0 \equiv - N(0) V_l^{({\rm eff})}$ with $V_l^{({\rm eff})}$ and $N(0)$ denoting 
the constant effective potential 
and the normal-state DOS per spin and unit volume at the Fermi level. 
Using the Green's function, we can also express the heat flux density ${\bm j}_{\rm Q}$ as
\begin{align}
{\bm j}_{\rm Q} = - \frac{N (0)}{2} \int_{- \infty}^\infty \varepsilon \langle {\bm v}_{\rm F} g^{\rm K} \rangle_{\rm F} d \varepsilon. \label{jQ}
\end{align}

Now, we use the following relations: 
\begin{subequations}
\begin{align}
&i \hbar {\bm v}_{\rm F} \cdot {\bm \partial} \check{g} 
= i \hbar {\bm v}_{\rm F} \cdot {\bm \nabla} \check{g} 
+ \left[ e {\bm v}_{\rm F} \cdot {\bm A} \check{\tau}_3, \check{g} \right], \\
&\left[ e {\bm v}_{\rm F} \cdot {\bm A} \check{\tau}_3, \check{g} \right]
+ \frac{i \hbar}{2} e {\bm v}_{\rm F} \cdot {\bm E} \frac{\partial}{\partial \varepsilon} 
\left\{ \check{\tau}_3, \check{g} \right\}
= \left[ e {\bm v}_{\rm F} \cdot {\bm A} \check{\tau}_3, \check{g} \right]_\circ, 
\end{align}
\end{subequations}
to rewrite Eq. (\ref{AQCEq}) as
\begin{align}
&\left[ (\varepsilon + e {\bm v}_{\rm F} \cdot {\bm A}) \check{\tau}_3 - \check{\Delta} 
- \check{\sigma}, \check{g} \right]_\circ 
+ i \hbar {\bm v}_{\rm F} \cdot {\bm \nabla} \check{g} \notag \\
& \ \ \ + \frac{i \hbar}{2} e \left( {\bm v}_{\rm F} \times {\bm B} \right) \cdot \frac{\partial}{\partial {\bm p}_{\rm F}}
\left\{ \check{\tau}_3, \check{g} \right\} = \check{0}. 
\label{AQCEq-2}
\end{align}
Introducing matrix $\check{h}$ as \cite{Eschrig}
\begin{align}
\check{h} = \check{\Delta} + \check{\sigma} - e {\bm v}_{\rm F} \cdot {\bm A} \check{\tau}_3 = 
\begin{bmatrix}
\hat{h}^{\rm R} & \hat{h}^{\rm K} \\
\hat{0} & \hat{h}^{\rm A}
\end{bmatrix}, 
\end{align}
we obtain the equations for $\hat{g}^{\rm R,A,K}$ as
\begin{subequations}
\begin{align}
&\left[ \varepsilon \hat{\tau}_3 - \hat{h}^{\rm R,A}, \hat{g}^{\rm R,A} \right]_\circ 
+ i \hbar {\bm v}_{\rm F} \cdot {\bm \nabla} \hat{g}^{\rm R,A} \notag \\
& \ \ \ + \frac{i \hbar}{2} e \left( {\bm v}_{\rm F} \times {\bm B} \right) \cdot \frac{\partial}{\partial {\bm p}_{\rm F}}
\left\{ \hat{\tau}_3, \hat{g}^{\rm R,A} \right\} = \hat{0}, 
\end{align}
\begin{align}
&\left[ \varepsilon \hat{\tau}_3, \hat{g}^{\rm K} \right]_\circ 
- \hat{h}^{\rm R} \circ \hat{g}^{\rm K} - \hat{h}^{\rm K} \circ \hat{g}^{\rm A} + \hat{g}^{\rm R} \circ \hat{h}^{\rm K} + \hat{g}^{\rm K} \circ \hat{h}^{\rm A} \notag \\
& \ \ \ + i \hbar {\bm v}_{\rm F} \cdot {\bm \nabla} \hat{g}^{\rm K} 
+ \frac{i \hbar}{2} e \left( {\bm v}_{\rm F} \times {\bm B} \right) \cdot \frac{\partial}{\partial {\bm p}_{\rm F}}
\left\{ \hat{\tau}_3, \hat{g}^{\rm K} \right\} = \hat{0}.
\end{align}
\end{subequations}
%

\section{Thermal Hall Conductivity \label{sec:III}}
We consider the linear response $\hat{X} = \hat{X}^{\rm le} + \delta \hat{X}$ 
to the thermal gradient ${\bm \nabla} T$,  
where $\hat{X}^{\rm le}$ is the solution in local equilibrium 
and $\delta \hat{X}$ is a term of the first order in ${\bm \nabla} T$ \cite{Vorontsov,Graf}. 
We also neglect the electric field ${\bm E}$, 
since the quasiparticle current and supercurrent counterflow 
in order to keep the charge current zero \cite{Ginzbrug,Izawa07}, 
except the circulating supercurrent of a vortex, 
and the charge in a pinned vortex due to the Lorentz force \cite{Ueki16}
does not contribute to thermal conductivity. 
Assuming extreme type-II superconductors
as $\lambda_0 \gg \xi_0(0)$ with an isolated pinned vortex, 
we furthermore use the Doppler shift method for the quasiclassical equations \cite{Dahm}, 
where $\lambda_0$ denotes magnetic penetration depth 
defined by $\lambda_0 \equiv [\mu_0 N(0) e^2 \langle v_{\rm F}^2 \rangle_{\rm F}]^{-1/2}$, 
the coherence length $\xi_0(0)$ is defined by 
$\xi_0(0) \equiv \hbar \langle v_{\rm F} \rangle_{\rm F} / \Delta_0(0)$, 
$\Delta_0(0)$ denotes the gap amplitude in clean superconductors 
at zero temperature and zero magnetic field,
and $\mu_0$ is the vacuum permeability. 
Then we can neglect the spatial derivatives of the phase-transformed Green's functions, 
the vector potential terms. 
Hereafter, we remove the superscript ``le" from these equations.

\subsection{Local equilibrium} 
Equations for the retarded and advanced Green's functions $\hat{g}^{\rm R,A}$ in local equilibrium are written as 
\begin{align}
&\left[ \varepsilon \hat{\tau}_3 -\hat{\Delta} - \hat{\sigma}^{\rm R,A}, \hat{g}^{\rm R,A} \right] 
+ i \hbar {\bm v}_{\rm F} \cdot {\bm \nabla} \hat{g}^{\rm R,A} \notag \\
& \ \ \ + \frac{i \hbar}{2} e \left( {\bm v}_{\rm F} \times {\bm B} \right) \cdot \frac{\partial}{\partial {\bm p}_{\rm F}}
\left\{ \hat{\tau}_3, \hat{g}^{\rm R,A} \right\}  = \hat{0}, \label{gRA}
\end{align}
where notation $[a, b]$ is defined by $[a, b] \equiv a b - b a$. 
We first expand formally in the quasiclassical parameter 
$\delta \equiv \hbar / \langle p_{\rm F} \rangle_{\rm F} \xi_0(0)$ 
as $g^{\rm R, A} = g_0^{\rm R, A} + g_1^{\rm R, A} + \cdots$, 
$f^{\rm R, A} = f_0^{\rm R, A} + f_1^{\rm R, A} + \cdots$, 
and $\Delta = \Delta_0 + \Delta_1 + \cdots$ \cite{Kita09}.  
We next take the direction of the magnetic field as ${\bm B} = B \hat{\bm z}$, 
and transform $f_0^{\rm R, A}$ and $\Delta_0$ 
as $f_0^{\rm R, A} = \tilde{f}_0^{\rm R, A} {\rm e}^{- i \varphi}$ and $\Delta_0 = \tilde{\Delta}_0 {\rm e}^{- i \varphi}$
with the azimuth angle $\varphi$ in the cylindrical coordinate system $(\rho \cos \varphi, \rho \sin \varphi, z)$. 
Neglecting the spatial dependence of $\tilde{\Delta}_0$ and
the imaginary part in $g_0^{\rm R, A}$ and $\tilde{f}_0^{\rm R, A}$, 
then we obtain 
 \begin{subequations}
\begin{align}
g_0^{\rm R, A} &= 
\frac{- i \varepsilon^{\rm R, A}}{\sqrt{\Delta^{\rm R,A}{}^2 - \varepsilon^{\rm R, A}{}^2}}
= \bar{g}_0^{\rm R, A}, \label{g0RA} \\
\tilde{f}_0^{\rm R, A} &= \frac{\Delta^{\rm R, A}}{\sqrt{\Delta^{\rm R,A}{}^2 - \varepsilon^{\rm R, A}{}^2}} = \bar{\tilde{f}}_0^{\rm R, A}. \label{f0RA}
\end{align} \label{g0RAf0RA}%
\end{subequations}
$\varepsilon^{\rm R, A}$ and $\Delta^{\rm R, A}$ are defined by 
\begin{subequations}
\begin{align}
\varepsilon^{\rm R} 
&\equiv \tilde{\varepsilon} + i \eta + \frac{i \hbar}{2 \tau} \langle g_0^{\rm R} \rangle_{\rm F}, \\
\varepsilon^{\rm A} 
&\equiv \tilde{\varepsilon} - i \eta + \frac{i \hbar}{2 \tau} \langle g_0^{\rm A} \rangle_{\rm F}, \\
\Delta^{\rm R, A} &\equiv |\tilde{\Delta}_0| \phi + \frac{\hbar}{2 \tau} \langle \tilde{f}_0^{\rm R, A} \rangle_{\rm F},  
\end{align} \label{epsRA}%
\end{subequations}
where $\eta$ denotes an infinitesimal positive constant and $\tilde{\varepsilon}$ is defined by  
\begin{align}
\tilde{\varepsilon} \equiv \varepsilon - m {\bm v}_{\rm F} \cdot {\bm v}_{\rm s}, 
\end{align}
with $m$ denoting the electron mass and ${\bm v}_{\rm s}$ denoting the superfluid velocity defined by 
\begin{align}
{\bm v}_{\rm s} \equiv - \frac{\hbar}{2m} {\bm \nabla} \varphi. 
\end{align}

Solving the equations for $g_1^{\rm R, A}$ and $f_1^{\rm R, A}$ 
in the almost same manner as Ref. \onlinecite{Kita09}, 
we also obtain 
\begin{subequations}
\begin{align}
{\bm v}_{\rm F} \cdot {\bm \nabla} g_1^{\rm R, A} 
&= - e ({\bm v}_{\rm F} \times {\bm B}) \cdot \frac{\partial g_0^{\rm R, A}}{\partial {\bm p}_{\rm F}}, 
\\ 
g_1^{\rm R, A} &= - \bar{g}_1^{\rm R, A}, \label{g1RA} \\ 
f_1^{\rm R, A} &= 0. \label{f1RA}
\end{align} \label{g1RAf1RA}%
\end{subequations}
%

\subsection{First-order response} 
Equations for the retarded and advanced Green's functions $\delta \hat{g}^{\rm R,A}$ 
in the first-order response are written as 
\begin{align}
&\left[ \varepsilon \hat{\tau}_3 - \hat{h}^{\rm R,A}, \delta \hat{g}^{\rm R,A} \right]_\circ 
- \left[ \delta \hat{h}^{\rm R,A}, \hat{g}^{\rm R,A} \right]_\circ 
+ i \hbar {\bm v}_{\rm F} \cdot {\bm \nabla} \delta \hat{g}^{\rm R,A} \notag \\
& \ \ \ + \frac{i \hbar}{2} e \left( {\bm v}_{\rm F} \times {\bm B} \right) \cdot \frac{\partial}{\partial {\bm p}_{\rm F}}
\left\{ \hat{\tau}_3, \delta \hat{g}^{\rm R,A} \right\} \notag \\
& \ \ \ + \frac{i \hbar}{2} e \left( {\bm v}_{\rm F} \times \delta {\bm B} \right) \cdot \frac{\partial}{\partial {\bm p}_{\rm F}}
\left\{ \hat{\tau}_3, \hat{g}^{\rm R,A} \right\} 
= \hat{0}, \label{deltagRAeq}
\end{align}
and equations for the Keldysh Green's functions $\delta \hat{g}^{\rm K}$ 
in the first-order response are given by 
\begin{align}
&\left[ \varepsilon \hat{\tau}_3, \delta \hat{g}^{\rm K} \right]_\circ 
+ i \hbar {\bm v}_{\rm F} \cdot {\bm \nabla} \hat{g}^{\rm K}
+ i \hbar {\bm v}_{\rm F} \cdot {\bm \nabla} \delta \hat{g}^{\rm K} \notag \\
& \ \ \ - \hat{h}^{\rm R} \circ \delta \hat{g}^{\rm K} - \delta \hat{h}^{\rm R} \circ \hat{g}^{\rm K}
- \hat{h}^{\rm K} \circ \delta \hat{g}^{\rm A} - \delta \hat{h}^{\rm K} \circ \hat{g}^{\rm A} \notag \\
& \ \ \ + \delta \hat{g}^{\rm R} \circ \hat{h}^{\rm K} + \hat{g}^{\rm R} \circ \delta \hat{h}^{\rm K}
+ \delta \hat{g}^{\rm K} \circ \hat{h}^{\rm A} + \hat{g}^{\rm K} \circ \delta \hat{h}^{\rm A} \notag \\
& \ \ \ + \frac{i \hbar}{2} e \left( {\bm v}_{\rm F} \times {\bm B} \right) \cdot \frac{\partial}{\partial {\bm p}_{\rm F}}
\left\{ \hat{\tau}_3, \delta \hat{g}^{\rm K} \right\} \notag \\
& \ \ \ + \frac{i \hbar}{2} e \left( {\bm v}_{\rm F} \times \delta {\bm B} \right) \cdot \frac{\partial}{\partial {\bm p}_{\rm F}}
\left\{ \hat{\tau}_3, \hat{g}^{\rm K} \right\} 
= \hat{0}. \label{deltagKeq}
\end{align}
Note that operator ${\bm \nabla}$ to $\hat{g}^{\rm K}$ and $\delta \hat{g}^{\rm K}$ only affects temperature in the distribution functions and the Green's functions except the distribution functions, respectively \cite{Vorontsov}.
Let us introduce $\delta \hat{g}^{\rm a}$ as \cite{Graf,Eschrig} 
\begin{align}
\delta \hat{g}^{\rm K} = \delta \hat{g}^{\rm R} \circ \tanh \frac{\varepsilon}{2 k_{\rm B} T} 
- \tanh \frac{\varepsilon}{2 k_{\rm B} T} \circ \delta \hat{g}^{\rm A} 
+ \delta \hat{g}^{\rm a}, \label{deltaga}
\end{align}
with
\begin{align}
\delta \hat{g}^{\rm a} =
\begin{bmatrix}
\delta g^{\rm a} & - i \delta f^{\rm a} \\
- i \delta \bar{f}^{\rm a} & \delta \bar{g}^{\rm a} 
\end{bmatrix}. 
\end{align}
The Green's functions $\delta g^{\rm R, A}$ are used to calculate the supercurrent, 
and $\delta g^{\rm a}$ is related to the quasiparticle current. 
Thus, we derive equation for $\delta g^{\rm a}$ to obtain the heat current. 
Substituting Eq. (\ref{deltaga}) into Eq. (\ref{deltagKeq}), 
and using $\hat{g}^{\rm K} = (\hat{g}^{\rm R} - \hat{g}^{\rm A}) \tanh \varepsilon / 2 k_{\rm B} T$ and 
Eq. (\ref{deltagRAeq}), the equation for $\delta \hat{g}^{\rm a}$ can be expressed as
\begin{align}
&\left[ \varepsilon \hat{\tau}_3, \delta \hat{g}^{\rm a} \right]_\circ 
+ i \hbar {\bm v}_{\rm F} \cdot {\bm \nabla} \delta \hat{g}^{\rm a} \notag \\
& \ \ \ - \hat{h}^{\rm R} \circ \delta \hat{g}^{\rm a} + \delta \hat{g}^{\rm a} \circ \hat{h}^{\rm A}
+ \hat{g}^{\rm R} \circ \delta \hat{h}^{\rm a} - \delta \hat{h}^{\rm a} \circ \hat{g}^{\rm A} \notag \\
& \ \ \ + i \hbar (\hat{g}^{\rm R} - \hat{g}^{\rm A}) {\bm v}_{\rm F} \cdot {\bm \nabla} \tanh \frac{\varepsilon}{2 k_{\rm B} T} \notag \\
& \ \ \ + \frac{i \hbar}{2} e \left( {\bm v}_{\rm F} \times {\bm B} \right) \cdot \frac{\partial}{\partial {\bm p}_{\rm F}}
\left\{ \hat{\tau}_3, \delta \hat{g}^{\rm a} \right\} 
= \hat{0}. \label{deltagaeq}
\end{align}
We also neglect the time derivative terms and the self-energy correction terms 
as $\langle \delta g^{\rm a} \rangle_{\rm F}$ and $\langle \delta f^{\rm a} \rangle_{\rm F}$ \cite{Graf}. 
Then the equations for $ \delta g^{\rm a}$ and $ \delta f^{\rm a}$ 
are obtained from the $(1,1)$ and $(1, 2)$ components in Eq. (\ref{deltagaeq}) as 
\begin{subequations}
\begin{align}
&\hbar {\bm v}_{\rm F} \cdot {\bm \nabla} \delta g^{\rm a} + 
\hbar (g^{\rm R} - g^{\rm A}) {\bm v}_{\rm F} \cdot {\bm \nabla} \tanh \frac{\varepsilon}{2 k_{\rm B} T} \notag \\
& \ \ \ + \hbar e ({\bm v}_{\rm F} \times {\bm B}) \cdot \frac{\partial \delta g^{\rm a}}{\partial {\bm p}_{\rm F}} 
- \Delta \phi \delta \bar{f}^{\rm a} - \Delta^* \phi^* \delta f^{\rm a} \notag \\
& \ \ \ + \frac{\hbar}{2 \tau} \langle g^{\rm R} - g^{\rm A} \rangle_{\rm F} \delta g^{\rm a} 
- \frac{\hbar}{2 \tau} \langle f^{\rm R} \rangle_{\rm F} \delta \bar{f}^{\rm a} 
- \frac{\hbar}{2 \tau} \langle \bar{f}^{\rm A} \rangle_{\rm F} \delta f^{\rm a} \notag \\ 
& \ \ \ = 0, \label{dgaeq} \\
&-2i \varepsilon \delta f^{\rm a} + \hbar {\bm v}_{\rm F} \cdot {\bm \nabla} \delta f^{\rm a} 
+ \hbar (f^{\rm R} - f^{\rm A}) {\bm v}_{\rm F} \cdot {\bm \nabla} \tanh \frac{\varepsilon}{2 k_{\rm B} T} \notag \\ 
& \ \ \ + \Delta \phi \delta \bar{g}^{\rm a} - \Delta \phi \delta g^{\rm a} 
+ \frac{\hbar}{2 \tau} \langle g^{\rm R} + \bar{g}^{\rm A} \rangle_{\rm F} \delta f^{\rm a} \notag \\
& \ \ \ + \frac{\hbar}{2 \tau} \langle f^{\rm R} \rangle_{\rm F} \delta \bar{g}^{\rm a} 
- \frac{\hbar}{2 \tau} \langle f^{\rm A} \rangle_{\rm F} \delta g^{\rm a} = 0. \label{dfaeq}
\end{align} \label{dgadfaeq}%
\end{subequations}

We first calculate the zeroth-order quantity $\delta g_0^{\rm a} + \delta \bar{g}_0^{\rm a}$ 
in $\delta$, 
since the heat current [Eq. (\ref{jQ})] can be rewritten as 
\begin{align}
{\bm j}_{\rm Q} = - \frac{N (0)}{4} \int_{- \infty}^\infty \varepsilon 
\langle {\bm v}_{\rm F} (\delta g^{\rm a} + \delta \bar{g}^{\rm a}) \rangle_{\rm F} d \varepsilon. \label{jQ2}
\end{align}
Transforming the zeroth-order quantity $\delta f_0^{\rm a}$ in $\delta$ 
as $\delta f_0^{\rm a} = \delta \tilde{f}_0^{\rm a} {\rm e}^{- i \varphi}$,
and neglecting the spatial derivatives of $\delta g_0^{\rm a}$, 
the equation for the zeroth-order quantity $\delta g_0^{\rm a}$ in $\delta$ 
is obtained from Eq. (\ref{dgaeq}) as 
\begin{subequations}
\begin{align}
&\hbar (g_0^{\rm R} - g_0^{\rm A}) {\bm v}_{\rm F} 
\cdot {\bm \nabla} \tanh \frac{\varepsilon}{2 k_{\rm B} T} 
+ \frac{\hbar}{2 \tau} \langle g_0^{\rm R} - g_0^{\rm A} \rangle_{\rm F} \delta g_0^{\rm a}
\notag \\
&- \left( |\tilde{\Delta}_0| \phi + \frac{\hbar}{2 \tau} \langle \tilde{f}_0^{\rm R} \rangle_{\rm F} \right) 
\delta \bar{\tilde{f}}_0^{\rm a} 
- \left( |\tilde{\Delta}_0| \phi^* + \frac{\hbar}{2 \tau} \langle \bar{\tilde{f}}_0^{\rm A} \rangle_{\rm F} \right) 
\delta \tilde{f}_0^{\rm a} \notag \\
&= 0,  \label{dg0aeq}
\end{align} 
and we obtain $\delta \tilde{f}_0^{\rm a}$ from the normalization condition \cite{Graf,Eschrig} as 
\begin{align}
\delta \tilde{f}_0^{\rm a} = - \frac{\tilde{f}_0^{\rm R} \delta \bar{g}_0^{\rm a} + \tilde{f}_0^{\rm A} \delta g_0^{\rm a}}{g_0^{\rm R} - \bar{g}_0^{\rm A}}. \label{df0aeq}
\end{align} \label{dg0adf0aeq}%
\end{subequations}
We can obtain the equation for $\delta \bar{g}_0^{\rm a}$ 
from the barred Eq. (\ref{dg0aeq})
and $\delta \bar{\tilde{f}}_0^{\rm a}$ from the barred Eq. (\ref{df0aeq}). 
We solve the equations for 
$\delta g_0^{\rm a}$, $\delta \bar{g}_0^{\rm a}$, 
$\delta \tilde{f}_0^{\rm a}$, and $\delta \bar{\tilde{f}}_0^{\rm a}$
choosing $\phi = \phi^*$ and using Eqs. (\ref{symG}) and (\ref{g0RAf0RA}). 
Then we obtain $\delta g^{\rm a} + \delta \bar{g}^{\rm a}$ as 
\begin{align}
&\delta g_0^{\rm a} + \delta \bar{g}_0^{\rm a} \notag \\
&= \frac{-4 {\rm Re} g_0^{{\rm R}2}}
{(\langle {\rm Re} g_0^{\rm R} \rangle_{\rm F} / \tau) {\rm Re} g_0^{\rm R} 
+ (2 |\tilde{\Delta}_0| \phi / \hbar + \langle {\rm Re} \tilde{f}_0^{\rm R} \rangle_{\rm F} / \tau) 
{\rm Re} \tilde{f}_0^{\rm R}} \notag \\
& \ \ \ \times {\bm v}_{\rm F} \cdot {\bm \nabla} \tanh \frac{\varepsilon}{2 k_{\rm B} T}. \label{dga0}
\end{align} 

We next calculate the first-order quantity $\delta g_1^{\rm a} + \delta \bar{g}_1^{\rm a}$ 
in $\delta$. 
Transforming the first-order quantity $\delta f_1^{\rm a}$ in $\delta$
as $\delta f_1^{\rm a} = \delta \tilde{f}_1^{\rm a} {\rm e}^{- i \varphi}$,
neglecting the spatial derivatives of $\delta g_1^{\rm a}$ and $\delta \tilde{f}_1^{\rm a}$, 
and using Eq. (\ref{f1RA}),
the equations for the first-order quantities $\delta g_1^{\rm a}$ and $\delta f_1^{\rm a}$ in $\delta$ 
are obtained from Eq. (\ref{dgadfaeq}) as
\begin{subequations}
\begin{align}
&\hbar (g_1^{\rm R} - g_1^{\rm A}) {\bm v}_{\rm F} \cdot {\bm \nabla} \tanh \frac{\varepsilon}{2 k_{\rm B} T}
+ \hbar e ({\bm v}_{\rm F} \times {\bm B}) \cdot \frac{\partial \delta g_0^{\rm a}}{\partial {\bm p}_{\rm F}} \notag \\
&- \left( |\tilde{\Delta}_0| \phi + \frac{\hbar}{2 \tau} \langle \tilde{f}_0^{\rm R} \rangle_{\rm F} \right) 
\delta \bar{\tilde{f}}_1^{\rm a} 
- \left( |\tilde{\Delta}_0| \phi^* + \frac{\hbar}{2 \tau} \langle \bar{\tilde{f}}_0^{\rm A} \rangle_{\rm F} \right)
\delta \tilde{f}_1^{\rm a} \notag \\
&+ \frac{\hbar}{2 \tau} \langle g_0^{\rm R} - g_0^{\rm A} \rangle_{\rm F} \delta g_1^{\rm a} 
+ \frac{\hbar}{2 \tau} \langle g_1^{\rm R} - g_1^{\rm A} \rangle_{\rm F} \delta g_0^{\rm a} = 0, \label{dga1eq} \\
&-2i \tilde{\varepsilon} \delta \tilde{f}_1^{\rm a} 
+ \frac{\hbar}{2 \tau} \langle g_1^{\rm R} + \bar{g}_1^{\rm A} \rangle_{\rm F} 
\delta \tilde{f}_0^{\rm a} 
+ \frac{\hbar}{2 \tau} \langle g_0^{\rm R} + \bar{g}_0^{\rm A} \rangle_{\rm F} 
\delta \tilde{f}_1^{\rm a}  \notag \\
&+ \left( |\tilde{\Delta}_0| \phi + \frac{\hbar}{2 \tau} \langle \tilde{f}_0^{\rm R} \rangle_{\rm F} \right) 
\delta \bar{g}_1^{\rm a} 
- \left( |\tilde{\Delta}_0| \phi + \frac{\hbar}{2 \tau} \langle \tilde{f}_0^{\rm A} \rangle_{\rm F} \right)
\delta g_1^{\rm a} \notag \\
&= 0. \label{dfa1eq}
\end{align} 
\end{subequations}
We solve the equations for 
$\delta g_1^{\rm a}$, $\delta \bar{g}_1^{\rm a}$, 
$\delta \tilde{f}_1^{\rm a}$, and $\delta \bar{\tilde{f}}_1^{\rm a}$
choosing $\phi = \phi^*$ and using Eqs. (\ref{symG}), (\ref{g0RAf0RA}), and (\ref{g1RA}).
Then we also obtain $\delta g_1^{\rm a} + \delta \bar{g}_1^{\rm a}$ as 
\begin{widetext}
\begin{align}
\delta g_1^{\rm a} + \delta \bar{g}_1^{\rm a} =
\frac{- e ({\bm v}_{\rm F} \times {\bm B}) 
(- 2 \tilde{\varepsilon} / \hbar + \langle {\rm Im} g_0^{\rm R} \rangle_{\rm F} / \tau)}
{(- 2 \tilde{\varepsilon} / \hbar + \langle {\rm Im} g_0^{\rm R} \rangle_{\rm F} / \tau)
(\langle {\rm Re} g_0^{\rm R} \rangle_{\rm F} / \tau)
+ (2 |\tilde{\Delta}_0| \phi / \hbar + \langle {\rm Re} \tilde{f}_0^{\rm R} \rangle_{\rm F} / \tau)
(\langle {\rm Im} \tilde{f}_0^{\rm R} \rangle_{\rm F} / \tau)}
\cdot \frac{\partial (\delta g_0^{\rm a} + \delta \bar{g}_0^{\rm a})}{\partial {\bm p}_{\rm F}}. \label{dga1}
\end{align} 
\end{widetext}
%

\subsection{Coefficient of thermal conductivity} 
We finally calculate the thermal conductivity in extreme type-II superconductors
with an isolated pinned vortex. 
The thermal conductivity $\underline{\kappa}$ is given 
by taking the spatial average of 
the local thermal conductivity $\underline{\kappa}^{\rm loc} ({\bm r})$
defined by ${\bm j}_{\rm Q} = - \underline{\kappa}^{\rm loc} {\bm \nabla} T$. 
We thereby obtain the coefficient of thermal conductivity as
\begin{widetext}
\begin{align}
\underline{\kappa} &= \frac{N(0)}{2 k_{\rm B} T^2} \int_{- \infty}^\infty d \varepsilon 
\Bigg[ \left\langle \frac{ {\bm v}_{\rm F} {\rm Re}g_0^{{\rm R}2} {\bm v}_{\rm F}}
{\left( \langle {\rm Re}g_0^{\rm R} \rangle_{\rm F} / \tau \right) {\rm Re} g_0^{\rm R} 
+ (2 |\tilde{\Delta}_0| \phi / \hbar 
+ \langle {\rm Re} \tilde{f}_0^{\rm R} \rangle_{\rm F} / \tau) 
{\rm Re} \tilde{f}_0^{\rm R}} \right\rangle \notag \\
& \ \ \ - \Bigg\langle \frac{ {\bm v}_{\rm F} e ( {\bm v}_{\rm F} \times {\bm B} ) 
(- 2 \tilde{\varepsilon} / \hbar + \langle {\rm Im} g_0^{\rm R} \rangle_{\rm F} / \tau)}
{(- 2 \tilde{\varepsilon} / \hbar + \langle {\rm Im} g_0^{\rm R} \rangle_{\rm F} / \tau)
(\langle {\rm Re} g_0^{\rm R} \rangle_{\rm F} / \tau)
+ (2 |\tilde{\Delta}_0| \phi / \hbar + \langle {\rm Re} \tilde{f}_0^{\rm R} \rangle_{\rm F} / \tau)
(\langle {\rm Im} \tilde{f}_0^{\rm R} \rangle_{\rm F} / \tau)} \notag \\
& \ \ \ \ \ \ \ \ \ \ \cdot \frac{\partial}{\partial {\bm p}_{\rm F}} 
\frac{ {\rm Re}g_0^{{\rm R}2} {\bm v}_{\rm F}}
{\left( \langle {\rm Re}g_0^{\rm R} \rangle_{\rm F} / \tau \right) {\rm Re} g_0^{\rm R} 
+ (2 |\tilde{\Delta}_0| \phi / \hbar 
+ \langle {\rm Re} \tilde{f}_0^{\rm R} \rangle_{\rm F} / \tau) {\rm Re} \tilde{f}_0^{\rm R}}
\Bigg\rangle \Bigg] \varepsilon^2 {\rm sech}^2 \frac{\varepsilon}{2 k_{\rm B} T}, \label{kappa}
\end{align}
\end{widetext}
where $\langle \cdots \rangle$ denotes the Fermi surface and spatial average. 
The first line in Eq. (\ref{kappa}) for $B = 0$ (i.e., ${\bm v}_{\rm s}={\bm 0}$) 
is the same as the longitudinal thermal conductivity 
proposed in Ref. \onlinecite{Graf}, 
and the second and third lines are 
the newly derived thermal Hall conductivity.  
Neglecting the Doppler shift effect as ${\bm v}_{\rm s}={\bm 0}$, 
the expression for the thermal conductivity in $d$-wave superconductors can be rewritten as the following simpler form: 
\begin{widetext}
\begin{align}
&\underline{\kappa} = \frac{N(0)}{2 k_{\rm B} T^2} \int_{- \infty}^\infty d \varepsilon 
\Bigg[ \left\langle \frac{ {\bm v}_{\rm F} {\rm Re}g_0^{{\rm R}2} {\bm v}_{\rm F}}{\left( \langle {\rm Re}g_0^{\rm R} \rangle_{\rm F} / \tau \right) {\rm Re} g_0^{\rm R} + (2 \Delta_0 \phi / \hbar) {\rm Re} f_0^{\rm R}} \right\rangle_{\rm F} \notag \\
& \ \ - \left\langle \frac{ {\bm v}_{\rm F} e ( {\bm v}_{\rm F} \times {\bm B} ) }{\langle {\rm Re}g_0^{\rm R} \rangle_{\rm F}  / \tau } \cdot \frac{\partial}{\partial {\bm p}_{\rm F}} 
 \frac{ {\rm Re}g_0^{{\rm R}2} {\bm v}_{\rm F}}{\left( \langle {\rm Re}g_0^{\rm R} \rangle_{\rm F} / \tau \right) {\rm Re} g_0^{\rm R} + (2 \Delta_0 \phi / \hbar) {\rm Re} f_0^{\rm R}}
\right\rangle_{\rm F} \Bigg] \varepsilon^2 {\rm sech}^2 \frac{\varepsilon}{2 k_{\rm B} T}. \label{kappa-d}
\end{align}
\end{widetext}

We can use the impurity self-energy by the $t$-matrix approximation, 
changing $\tau$ in Eqs.  (\ref{epsRA}) and (\ref{kappa}) as \cite{Graf,Vorontsov,Kato00}
\begin{subequations}
\begin{align}
\frac{\langle  {\rm Re} g_0^{\rm R} \rangle_{\rm F}}{\tau} \to \frac{{\rm Re} \tau_0^{\rm R} + (\langle {\rm Im} g_0^{\rm R} \rangle_{\rm F} / \langle  {\rm Re} g_0^{\rm R} \rangle_{\rm F}) {\rm Im} \tau_0^{\rm R}}{| \tau_0^{\rm R} |^2} \langle  {\rm Re} g_0^{\rm R} \rangle_{\rm F}, \\
\frac{\langle  {\rm Im} g_0^{\rm R} \rangle_{\rm F}}{\tau} \to \frac{{\rm Re} \tau_0^{\rm R} - (\langle {\rm Re} g_0^{\rm R} \rangle_{\rm F} / \langle  {\rm Im} g_0^{\rm R} \rangle_{\rm F}) {\rm Im} \tau_0^{\rm R}}{| \tau_0^{\rm R} |^2} \langle  {\rm Im} g_0^{\rm R} \rangle_{\rm F}, \\
\frac{\langle  {\rm Re} \tilde{f}_0^{\rm R} \rangle_{\rm F}}{\tau} \to \frac{{\rm Re} \tau_0^{\rm R} + (\langle {\rm Im} \tilde{f}_0^{\rm R} \rangle_{\rm F} / \langle  {\rm Re} \tilde{f}_0^{\rm R} \rangle_{\rm F}) {\rm Im} \tau_0^{\rm R}}{| \tau_0^{\rm R} |^2} \langle  {\rm Re} \tilde{f}_0^{\rm R} \rangle_{\rm F}, 
\end{align}
\begin{align}
\frac{\langle  {\rm Im} \tilde{f}_0^{\rm R} \rangle_{\rm F}}{\tau} \to \frac{{\rm Re} \tau_0^{\rm R} - (\langle {\rm Re} \tilde{f}_0^{\rm R} \rangle_{\rm F} / \langle  {\rm Im} \tilde{f}_0^{\rm R} \rangle_{\rm F}) {\rm Im} \tau_0^{\rm R}}{| \tau_0^{\rm R} |^2} \langle  {\rm Im} \tilde{f}_0^{\rm R} \rangle_{\rm F}. 
\end{align}
\end{subequations}
$\tau_0^{\rm R}$ is defined by
\begin{align}
&\tau_0^{\rm R} \equiv \tau_0 \left[ \cos^2 \delta_0 + \sin^2 \delta_0 \left( \langle g_0^{\rm R} \rangle_{\rm F}^2 + \langle \tilde{f}_0^{\rm R} \rangle_{\rm F}^2 \right) \right],
\end{align}
where $\tau_0$ and $\delta_0$ are the relaxation time in the normal state and the scattering phase shift 
given by 
\begin{align}
\frac{\hbar}{\tau_0} = \frac{2 \pi n_a N(0) U^{{\rm imp}2}}{1 + \left[ \pi N(0) U^{\rm imp} \right]^2}, 
 \ \ \ \tan \delta_0 = - \pi N(0) U^{\rm imp}
\end{align}
with $n_a$ and $U^{\rm imp}$ denoting the density of impurities and the impurity potential.
Using $g_0^{\rm R} \to 1$ for the normal-state limit $\Delta \to 0$, 
we obtain the expression for the thermal conductivity in normal metal with a spherical Fermi surface as 
\begin{subequations}
\begin{align}
\kappa_{xx}^{(n)} &= \frac{\pi^2}{3} \left( \frac{k_{\rm B}}{e} \right)^2 \sigma_{xx}^{(n)} T, \\
\kappa_{xy}^{(n)} &= \frac{\tau_0 e B}{m} \kappa_{xx}^{(n)}, \label{kappaxyn}
\end{align}
\end{subequations}
where $\sigma_{xx}^{(n)} = \tau_0 e^2 n / m$ denotes the dc conductivity in normal state obtained from the Drude model,
$n = (2 / 3) m N (0) v_{\rm F}^2$ is the electron density.

\begin{figure}[t]
        \begin{center}
                \includegraphics[width=0.9\linewidth]{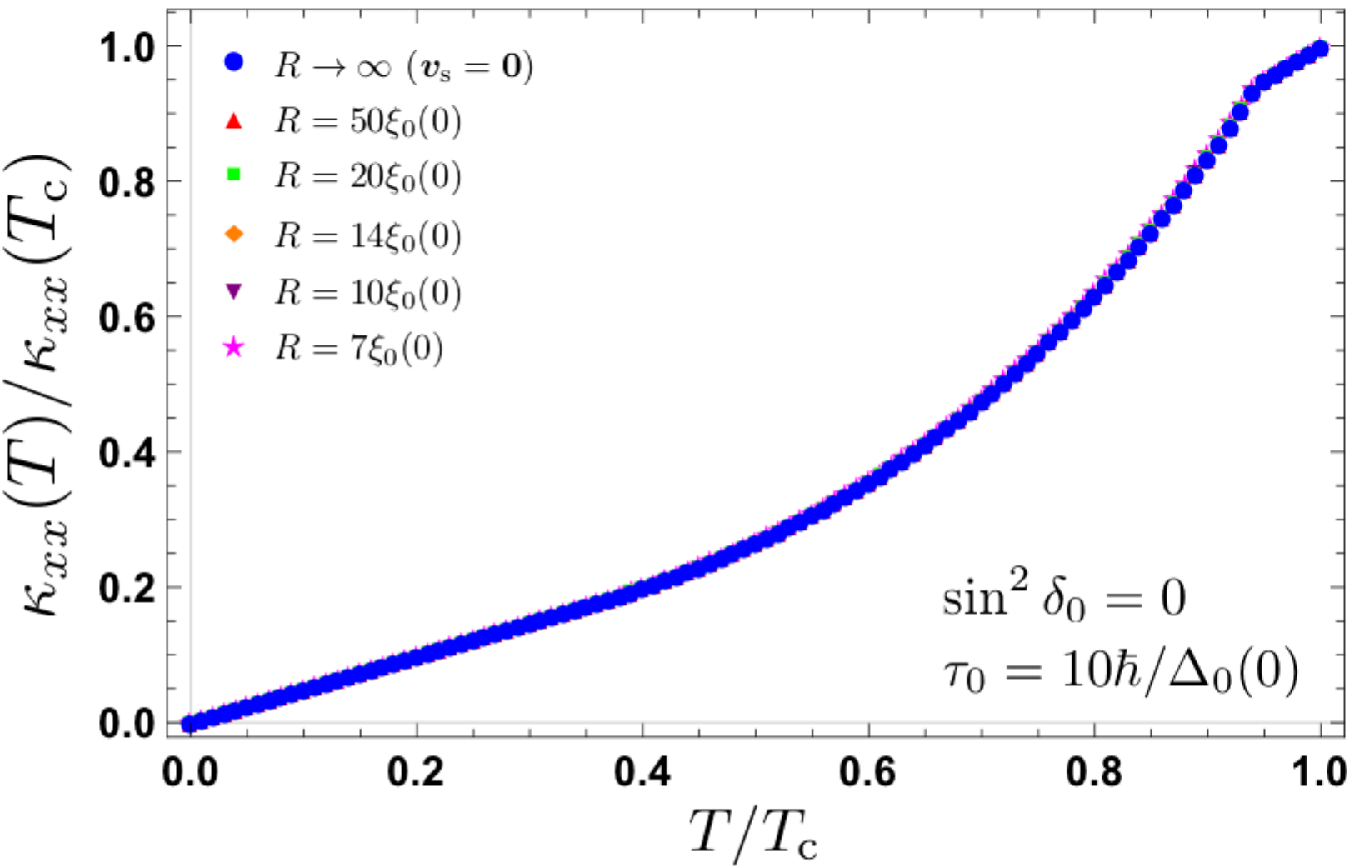}
                \end{center}
\caption{Temperature dependence of the longitudinal thermal conductivity $\kappa_{xx}$ 
normalized by the value at the transition temperature 
without impurities and external fields $T_{\rm c}$ 
for the different vortex radii $R = 7 \xi_0(0), 10 \xi_0(0), \dots, \infty$ 
at the relaxation time $\tau_0 = 10 \hbar / \Delta_0(0)$ 
and scattering phase shift $\sin^2 \delta_0 = 0$.}
\label{fig2}
\end{figure}
\begin{figure}[t]
        \begin{center}
                \includegraphics[width=0.9\linewidth]{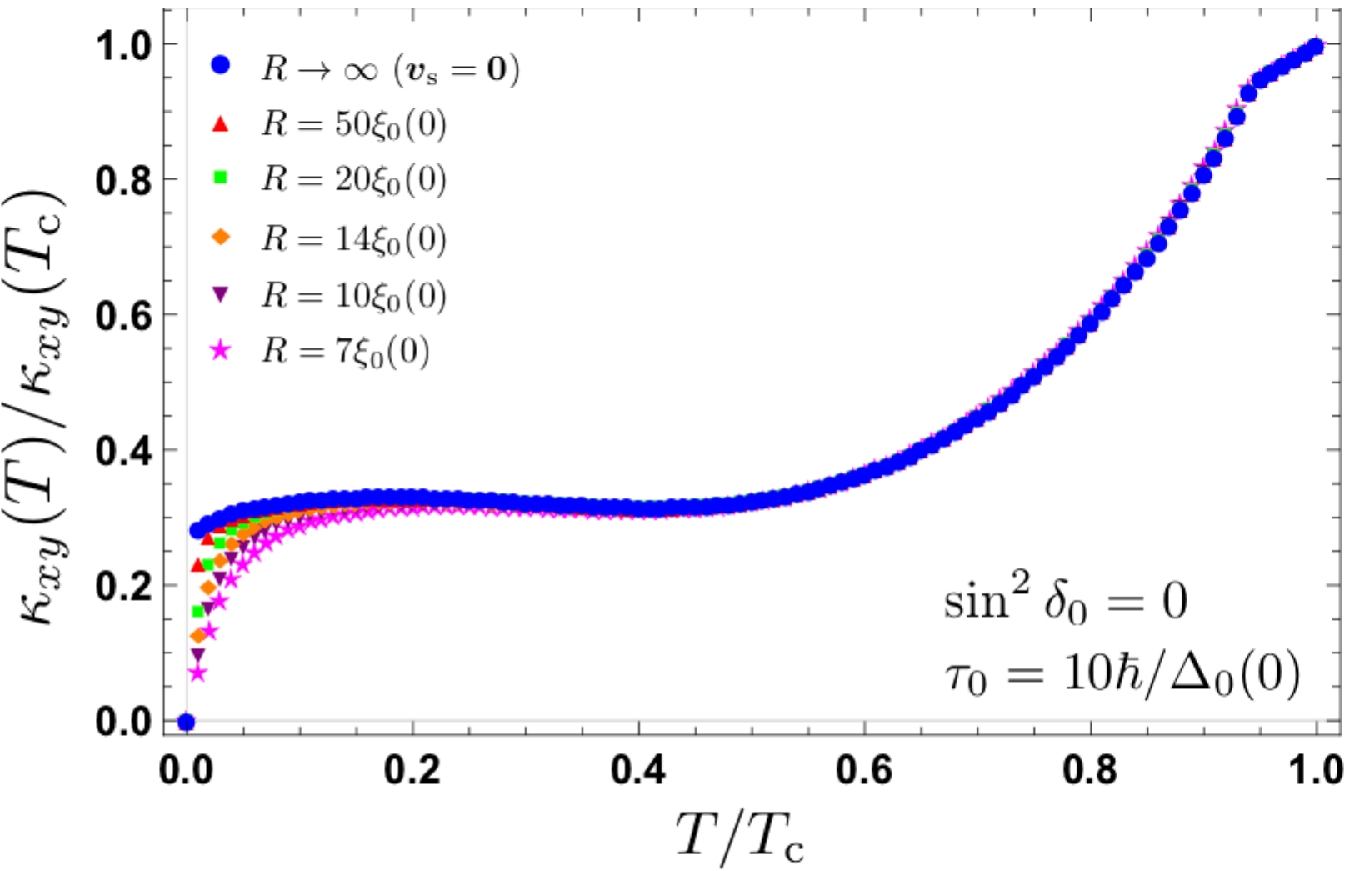}
                \end{center}
\caption{Temperature dependence of the thermal Hall conductivity $\kappa_{xy}$ 
normalized by the value at the transition temperature 
without impurities and external fields $T_{\rm c}$ 
for the different vortex radii $R = 7 \xi_0(0), 10 \xi_0(0), \dots, \infty$ 
at the relaxation time $\tau_0 = 10 \hbar / \Delta_0(0)$ 
and scattering phase shift $\sin^2 \delta_0 = 0$.}
\label{fig4}
\end{figure}
\begin{figure}[t]
        \begin{center}
                \includegraphics[width=0.9\linewidth]{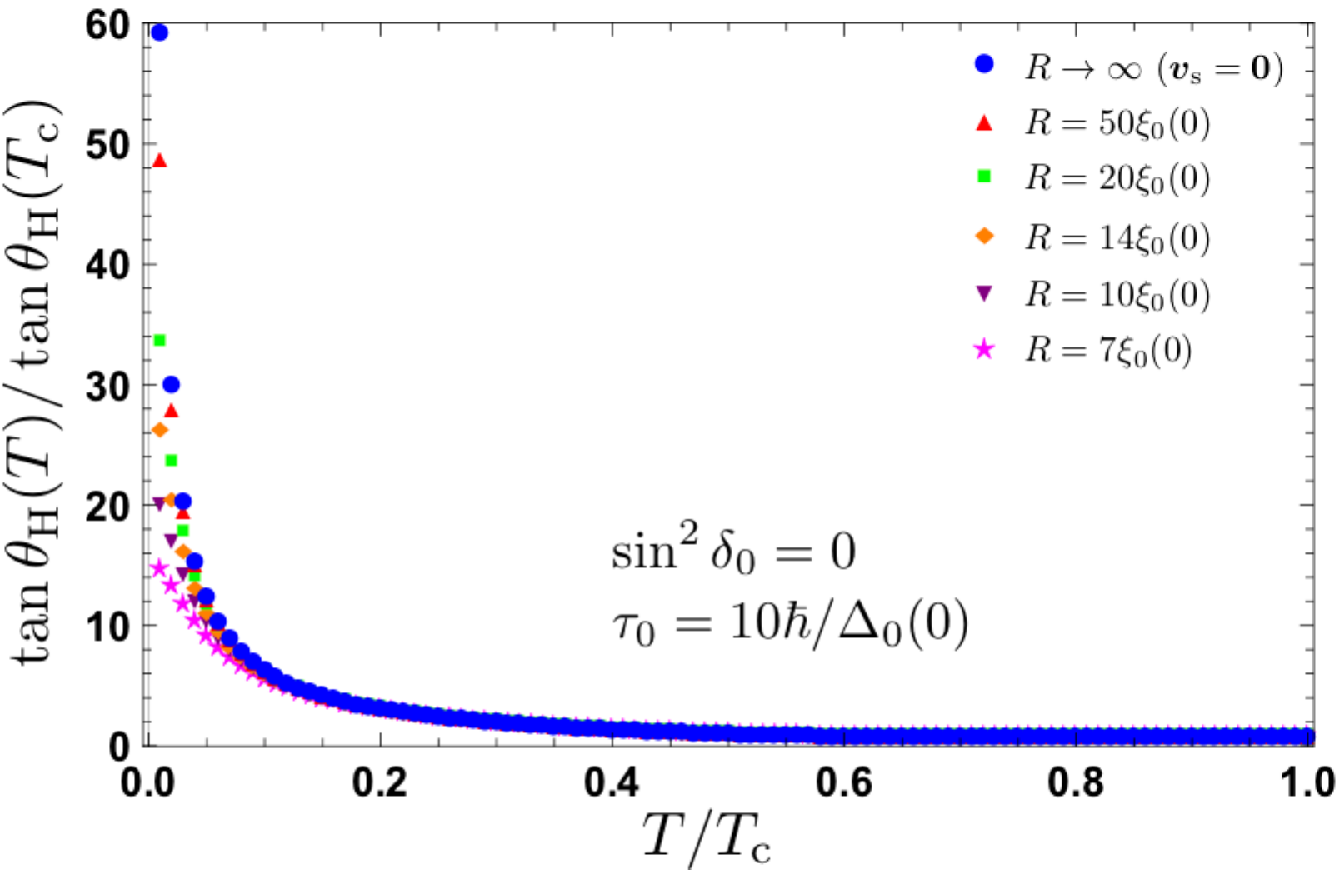}
                \end{center}
\caption{Temperature dependence of the thermal Hall angle $\tan \theta_{\rm H}$ 
normalized by the value at the transition temperature 
without impurities and external fields $T_{\rm c}$ 
for the different vortex radii $R = 7 \xi_0(0), 10 \xi_0(0), \dots, \infty$ 
at the relaxation time $\tau_0 = 10 \hbar / \Delta_0(0)$ 
and scattering phase shift $\sin^2 \delta_0 = 0$.}
\label{fig6}
\end{figure}
%

\section{Numerical Results \label{sec:IV}}
Here we present numerical results for the thermal Hall conductivity 
in $d$-wave superconductors with a cylindrical Fermi surface. 
Then $\langle \cdots \rangle$ is given by 
\begin{align}
\langle \cdots \rangle = \frac{1}{\pi R^2} \int_0^R d\rho \rho \int_0^{2 \pi} d\varphi \int_0^{2 \pi} \frac{d\varphi_{\bm p}}{2 \pi} \cdots,
\end{align}
where $\varphi_{\bm p}$ denotes the angle of the ${\bm p}$ vector in the two-dimensional polar coordinate system 
as ${\bm p} = (p \cos \varphi_{\bm p}, p \sin \varphi_{\bm p})$, 
and $R$ denotes the radius of the vortex lattice unit cell area 
given by $R \approx \xi_0(0) \sqrt{H_{\rm c2} / H}$ for $R \gg \xi_0(0)$
with $H_{c2}$ denoting the upper critical magnetic-field \cite{Kubert}. 
We consider extreme type-II superconductors with an isolated vortex, 
and assume that the magnetic field $B$ is a very small constant within $r \le R$.

To discuss the Volovik and impurity effects qualitatively, 
we first solve Eqs. (\ref{gapeq}) and (\ref{g0RAf0RA}) self-consistently to obtain 
$(g_0^{\rm R}, f_0^{\rm R}, \Delta_0)$, 
and substituting the resulting solutions into Eq. (\ref{kappa}) or (\ref{kappa-d}), 
we can obtain the thermal conductivity and the thermal Hall angle. 
Hereafter, we choose the cutoff energy 
in the gap equation [Eq. (\ref{gapeq})] as $\varepsilon_{\rm c} = 40 k_{\rm B} T_{\rm c}$, 
and fix the parameter to $\eta = 0.00001 \Delta_0(0)$, 
where $T_{\rm c}$ denotes 
the superconducting transition temperature without impurities and external fields. 
We also adopt a model $d$-wave pairing as 
$\phi = \sqrt{2} \cos 2 \varphi_{\bm p}$,
and the direction of applied thermal gradient is the same as the measurement 
in Ref. \onlinecite{Zhang01}.

\begin{figure}[t]
        \begin{center}
                \includegraphics[width=0.9\linewidth]{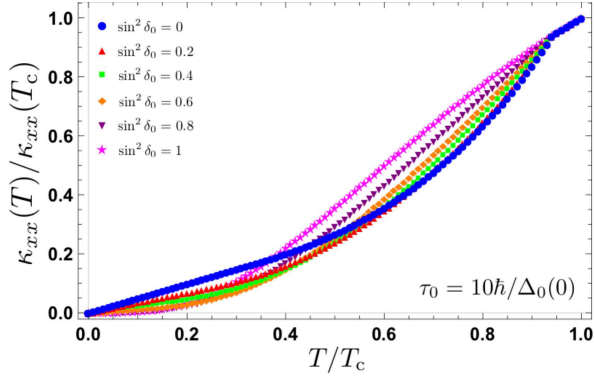}
                \end{center}
 \caption{Temperature dependence of the longitudinal thermal conductivity $\kappa_{xx}$ 
normalized by the value at the transition temperature 
without impurities and external fields $T_{\rm c}$ 
for the different scattering phase shifts $\sin^2 \delta_0 = 0, 0.2, \dots, 1$ 
at the relaxation time $\tau_0 = 10 \hbar / \Delta_0(0)$.}
\label{fig7}
\end{figure}
\begin{figure}[t]
        \begin{center}
                \includegraphics[width=0.9\linewidth]{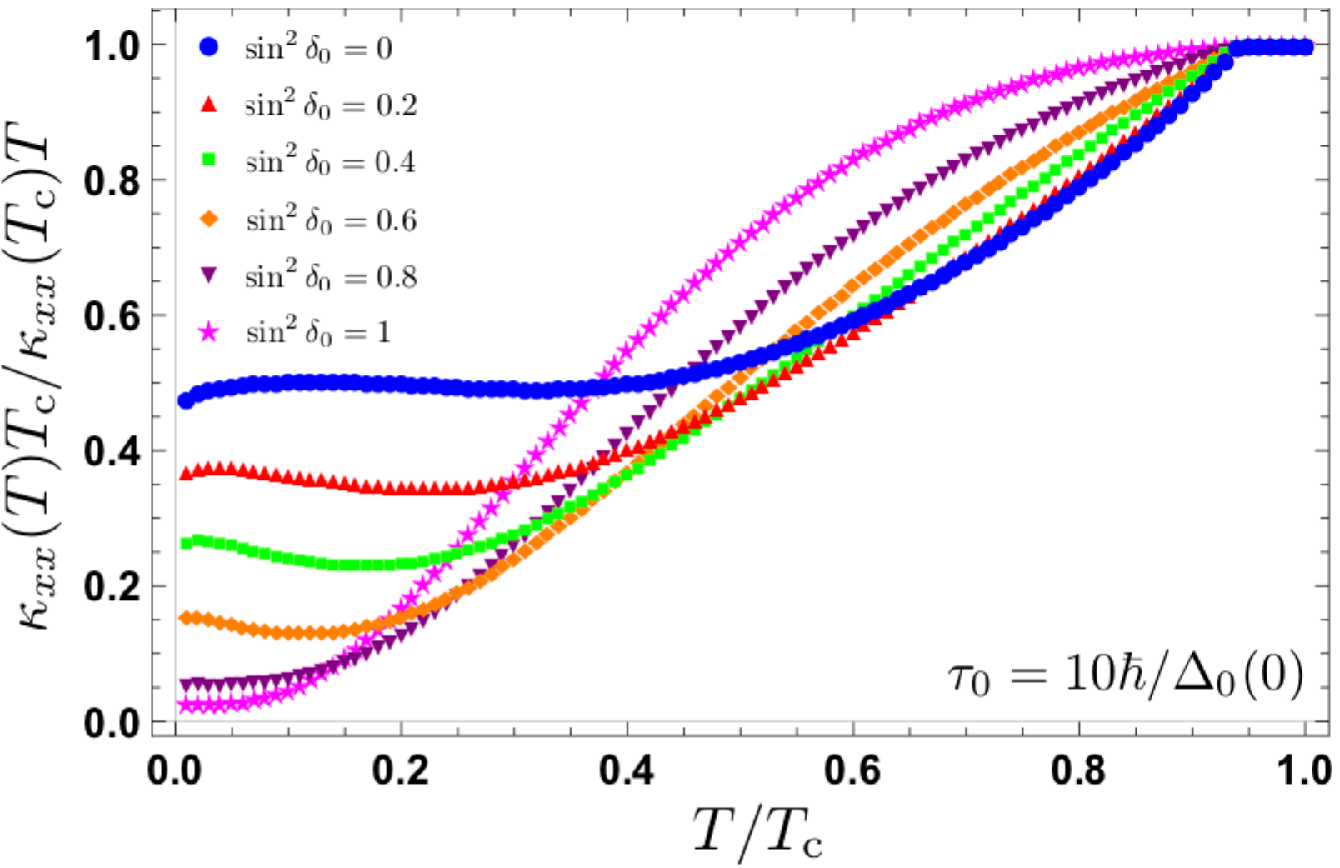}
                \end{center}
 \caption{Temperature dependence of the longitudinal thermal conductivity 
divided by temperature $\kappa_{xx} / T$ 
normalized by the value at the transition temperature 
without impurities and external fields $T_{\rm c}$ 
for the different scattering phase shifts $\sin^2 \delta_0 = 0, 0.2, \dots, 1$ 
at the relaxation time $\tau_0 = 10 \hbar / \Delta_0(0)$.}
\label{fig8}
\end{figure}
\begin{figure}[t]
        \begin{center}
                \includegraphics[width=0.9\linewidth]{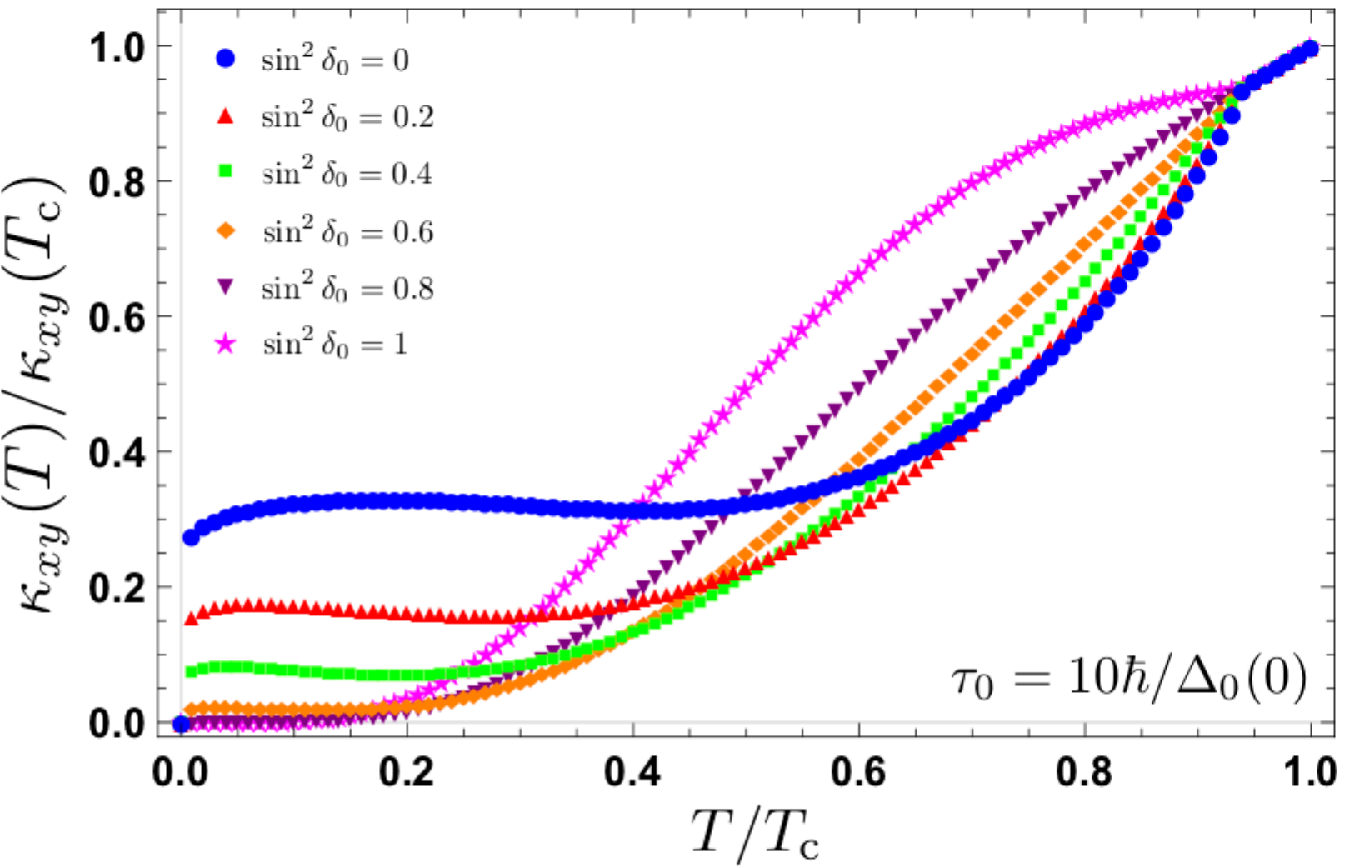}
                \end{center}
 \caption{Temperature dependence of the thermal Hall conductivity $\kappa_{xy}$ 
normalized by the value at the transition temperature 
without impurities and external fields $T_{\rm c}$ 
for the different scattering phase shifts $\sin^2 \delta_0 = 0, 0.2, \dots, 1$ 
at the normal relaxation time $\tau_0 = 10 \hbar / \Delta_0(0)$.}
\label{fig9}
\end{figure}
\begin{figure}[t]
        \begin{center}
                \includegraphics[width=0.9\linewidth]{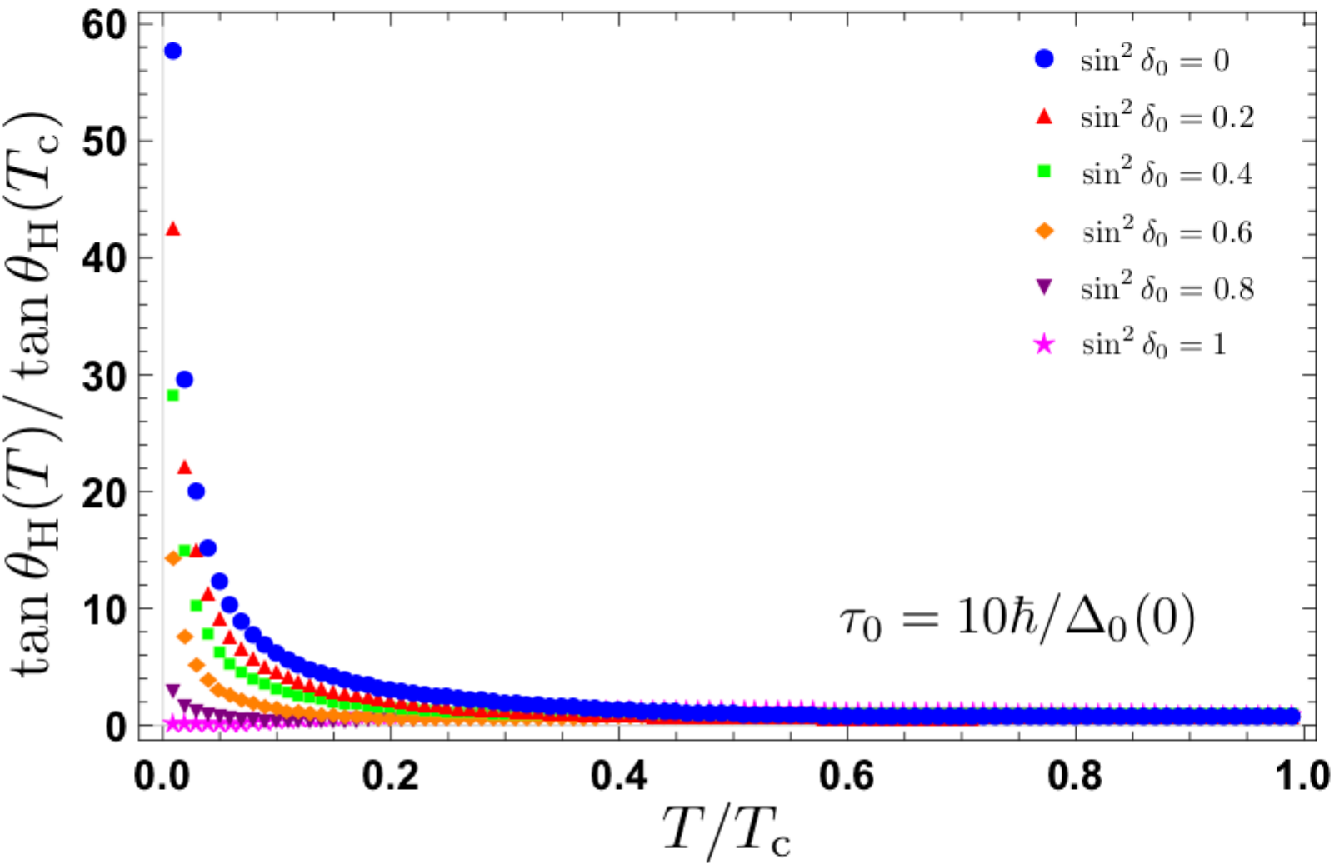}
                \end{center}
 \caption{Temperature dependence of the thermal Hall angle $\tan \theta_{\rm H}$
normalized by the value at the transition temperature 
without impurities and external fields $T_{\rm c}$ 
for the different scattering phase shifts $\sin^2 \delta_0 = 0, 0.2, \dots, 1$ 
at the normal relaxation time $\tau_0 = 10 \hbar / \Delta_0(0)$.}
\label{fig10}
\end{figure}
\begin{figure}[t]
        \begin{center}
                \includegraphics[width=0.9\linewidth]{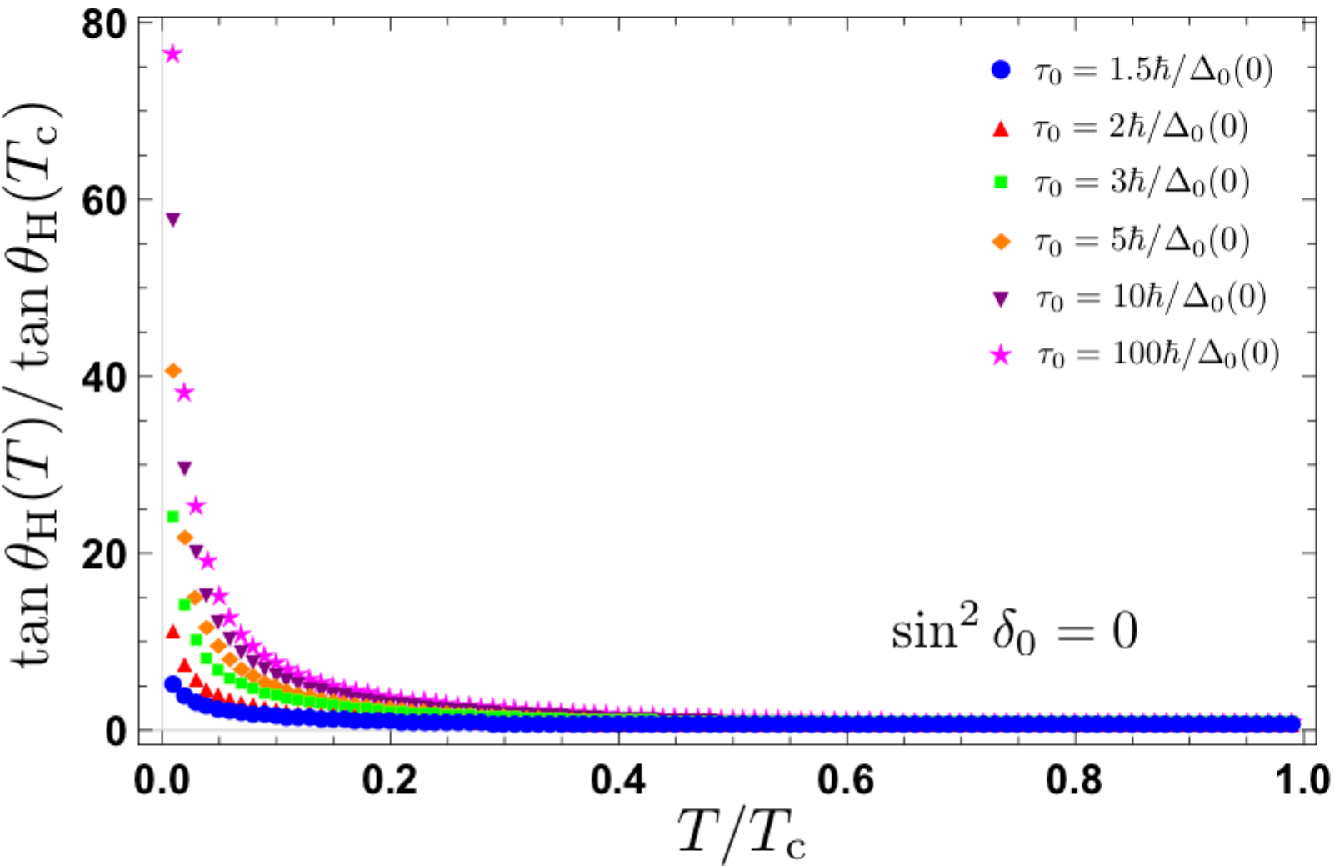}
                \end{center}
 \caption{Temperature dependence of the thermal Hall angle $\tan \theta_{\rm H}$
by the value at the transition temperature 
without impurities and external fields $T_{\rm c}$ 
for the different normal relaxation times 
$\tau_0 = 1.5 \hbar / \Delta_0(0)$, $2 \hbar / \Delta_0(0)$, \dots, 100 $\hbar / \Delta_0(0)$ 
in the Born limit as $\sin^2 \delta_0 = 0$.}
\label{fig11}
\end{figure}
\begin{figure}[t]
        \begin{center}
                \includegraphics[width=0.9\linewidth]{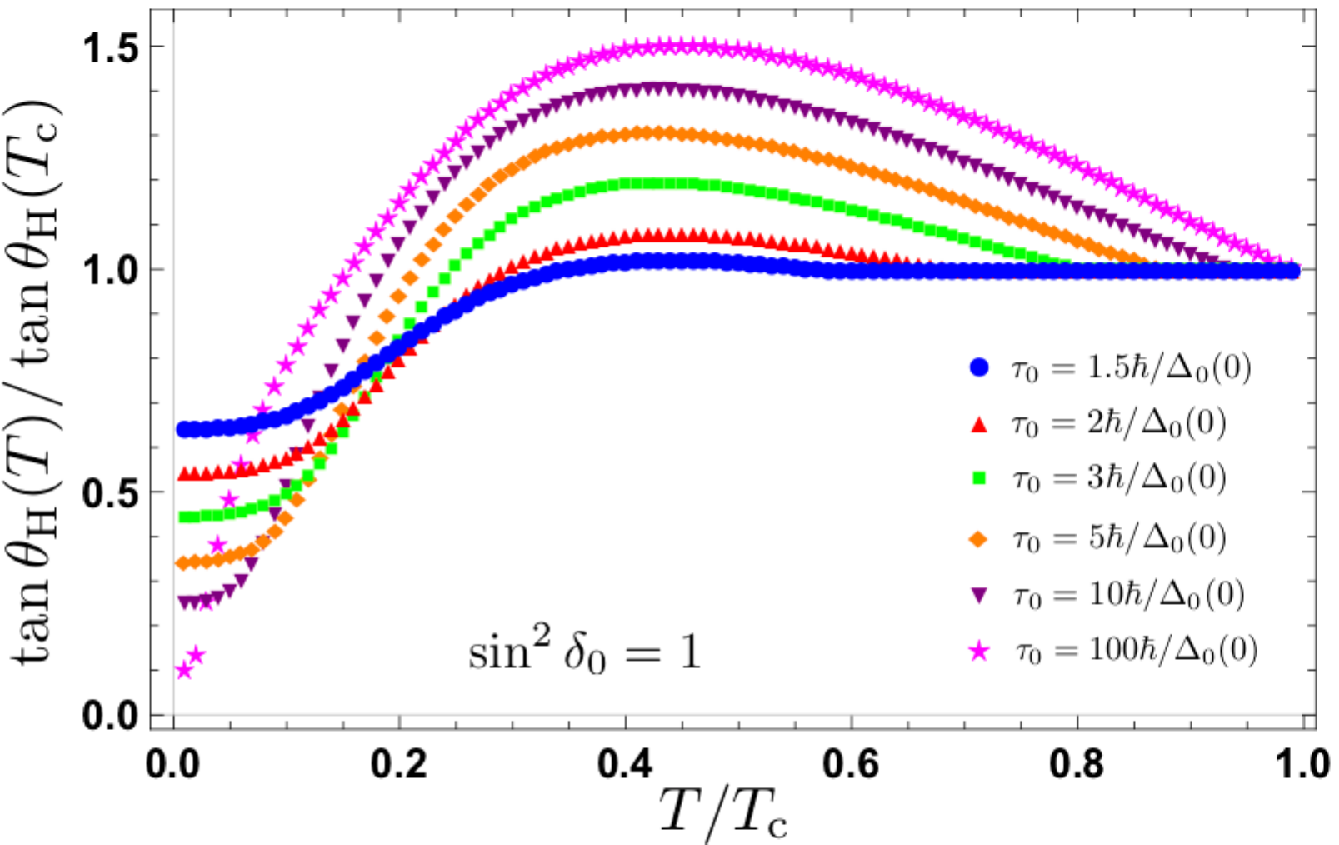}
                \end{center}
 \caption{Temperature dependence of the thermal Hall angle $\tan \theta_{\rm H}$
by the value at the transition temperature 
without impurities and external fields $T_{\rm c}$ 
for the different normal relaxation times 
$\tau_0 = 1.5 \hbar / \Delta_0(0)$, $2 \hbar / \Delta_0(0)$, \dots, 100 $\hbar / \Delta_0(0)$ 
in the unitarity limit as $\sin^2 \delta_0=1$.}
\label{fig12}
\end{figure}
\begin{figure}[t]
        \begin{center}
                \includegraphics[width=0.9\linewidth]{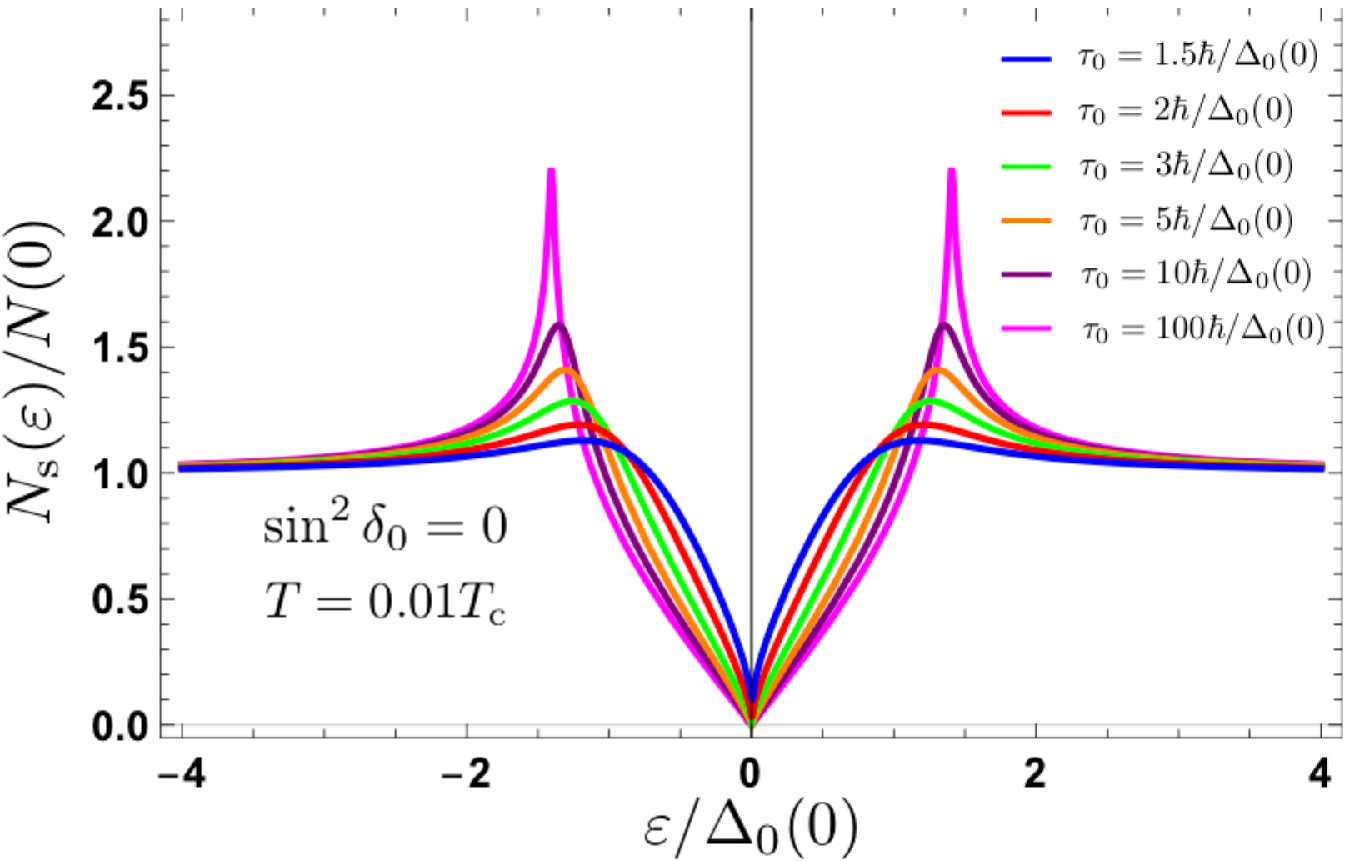}
                \end{center}
 \caption{Quasiparticle DOS $N_{\rm s} (\varepsilon)$ 
normalized by the normal DOS at the Fermi surface $N(0)$
for the different normal relaxation times 
$\tau_0 = 1.5 \hbar / \Delta_0(0)$, $2 \hbar / \Delta_0(0)$, \dots, 100 $\hbar / \Delta_0(0)$ 
in the Born limit as $\sin^2 \delta_0 = 0$ at $T = 0.01 T_{\rm c}$.}
\label{fig13}
\end{figure}
\begin{figure}[t]
        \begin{center}
                \includegraphics[width=0.9\linewidth]{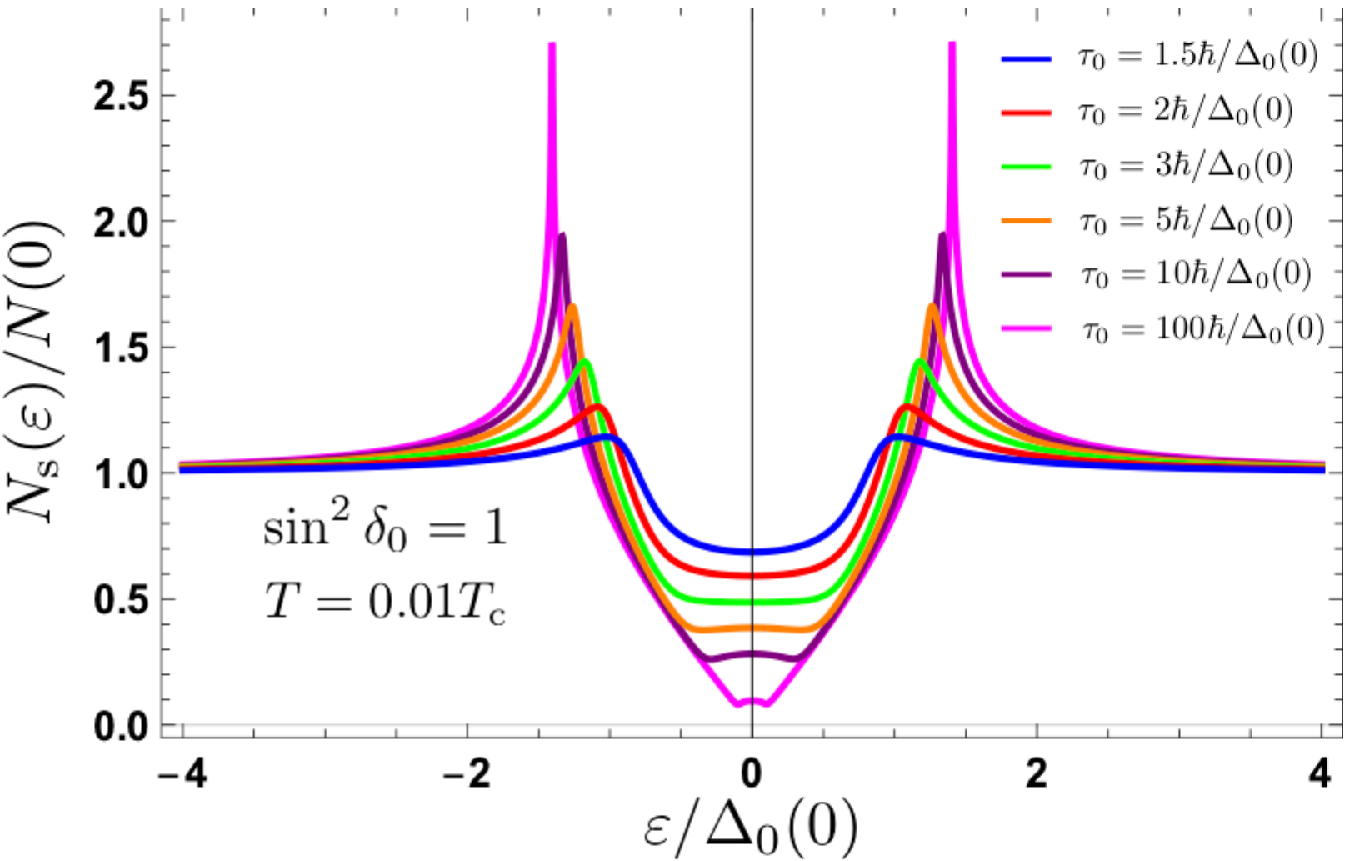}
                \end{center}
 \caption{Quasiparticle DOS $N_{\rm s} (\varepsilon)$ 
normalized by the normal DOS at the Fermi surface $N(0)$
for the different normal relaxation times 
$\tau_0 = 1.5 \hbar / \Delta_0(0)$, $2 \hbar / \Delta_0(0)$, \dots, 100 $\hbar / \Delta_0(0)$ 
in the unitarity limit as $\sin^2 \delta_0=1$ at $T = 0.01 T_{\rm c}$.}
\label{fig14}
\end{figure}
%

\subsection{Volovik effect} 
We first discuss the Doppler shift effect on 
the thermal conductivity and the thermal Hall angle in $d$-wave superconductors. 
It is shown that the Doppler shift effect can be neglected at very weak magnetic fields. 
In this subsection, we fix the parameters as $\sin^2 \delta_0 = 0$ (the Born-type impurity) 
and $\tau_0 = 10 \hbar / \Delta_0(0)$.

Figure \ref{fig2} plots the longitudinal thermal conductivity $\kappa_{xx}$ 
normalized by the value at the transition temperature without impurities and external fields 
$T_{\rm c}$, 
for different vortex radii $R = 7 \xi_0(0), 10 \xi_0(0), \dots, \infty$ as a function of temperature.
The thermal conductivity for $R \to \infty$ are the same as that for ${\bm v}_{\rm s} = {\bm 0}$. 
We observe that the normalized longitudinal thermal conductivity decreases 
as temperature decreases from $T = T_{\rm c}$, 
and $T$-linear behavior for $d$-wave superconductors at low temperatures.
These behaviors at low temperatures are well-known results \cite{Coffey,Graf}.

Figure \ref{fig4} plots the thermal Hall conductivity $\kappa_{xy}$ 
normalized by the value at the transition temperature 
without impurities and external fields $T_{\rm c}$, 
for the different vortex radii $R = 7 \xi_0(0), 10 \xi_0(0), \dots, \infty$ as a function of temperature.
We observe that the normalized thermal Hall conductivity 
varies more as a function of the vortex radius compared with the longitudinal component.
The normalized thermal Hall conductivity has a finite value even near $T=0$, 
and decreases significantly around $T=0.05T_{\rm c}$ 
as the vortex radius decreases from $R \to \infty$. 
We here note that 
the thermal Hall conductivity (without the normalization) is proportional to the magnetic field 
in the temperature range where the normalized thermal Hall conductivity 
does not change as a function of the vortex radius, 
since the thermal Hall conductivity at $T=T_{\rm c}$ used to normalize
is also proportional to the magnetic field as seen in Eq. (\ref{kappaxyn}). 
It means that the Doppler shift effect is very small in the temperature range 
where the normalized thermal Hall conductivity does not almost change.

Figure \ref{fig6} plots 
the thermal Hall angle $\tan \theta_{\rm H} \equiv \kappa_{xy} / \kappa_{xx}$ 
normalized by the value at the transition temperature 
without impurities and external fields $T_{\rm c}$, 
for the different vortex radii $R = 7 \xi_0(0), 10 \xi_0(0), \dots, \infty$ 
as a function of temperature.
We observe that the normalized thermal Hall angle greatly enhanced 
in $d$-wave superconductors.

We confirmed that the {normalized thermal conductivity and the normalized thermal Hall angle 
approach that for ${\bm v}_{\rm s} = {\bm 0}$ as the vortex radius $R$ increases, 
and the normalized thermal conductivity and the normalized thermal Hall angle 
for ${\bm v}_{\rm s}={\bm 0}$ are almost equal to that at $R = 1000 \xi_0(0)$ 
within $T \ge 0.01 T_{\rm c}$.

\subsection{Impurity effect} 
We finally discuss the impurity effect on the thermal conductivity and the thermal Hall angle 
normalized by the value at the transition temperature 
without impurities and external fields $T_{\rm c}$ 
in $d$-wave superconductors for ${\bm v}_{\rm s} = {\bm 0}$, 
since the experimental results on
the longitudinal thermal conductivity in heavy-fermion superconductors 
and Zn-doped YBCO were explained
by taking into account impurity scattering close to the unitarity limit 
\cite{Coffey,Hirschfeld96,Schmitt,Hirschfeld88},
and $\tan \theta_{\rm H} / H$ and $\kappa_{xy} / H$ in the zero magnetic-field limit $H \to 0$
were estimated experimentally by the initial slope in the $H$ dependence of $\kappa_{xy}$
in Refs. \onlinecite{Zhang01} and \onlinecite{Kasahara}. 
The thermal Hall conductivity and the thermal Hall angle at $T=T_{\rm c}$ used to normalize 
are proportional to the ``internal" magnetic field $B$ [see Eq. (\ref{kappaxyn})], 
and the internal magnetic field $B$ in extreme type-II superconductors also approaches zero 
as the external magnetic field $H$ approaches zero.
Thus, the normalized thermal Hall conductivity and the normalized thermal Hall angle 
in extreme type-II superconductors for ${\bm v}_{\rm s} = {\bm 0}$ 
are the same as evaluating 
$\lim_{H \to 0} \tan \theta_{\rm H} / H$ and $\lim_{H \to 0} \kappa_{xy} / H$.
Here we discuss the contribution 
from the scattering of quasiparticles on vortices \cite{Durst}. 
The thermal Hall conductivity due to the vortex scattering at weak magnetic fields is given by 
$\kappa_{xy}^{\rm vortex} \sim T \sqrt{H}$.
Since $\kappa_{xy}^{\rm vortex}/H$ diverges for $H \to 0$ 
regardless of temperature, 
and the longitudinal thermal conductivity is given by $\kappa_{xx} \sim T$, 
$\tan \theta_{\rm H} / H$, and $\kappa_{xy} / H$ for $H \to 0$ 
are dominated by the Lorentz force and the impurities. 
Therefore, we can study the impurity effect on thermally excited quasiparticles outside the core 
in extreme type-II superconductors, 
estimating 
$\lim_{H \to 0} \tan \theta_{\rm H} / H$ and $\lim_{H \to 0} \kappa_{xy} / H$ 
experimentally.

Figures \ref{fig7}, \ref{fig8}, \ref{fig9}, and \ref{fig10} plot 
the longitudinal thermal conductivity $\kappa_{xx} $ and $\kappa_{xx} / T$, 
the thermal Hall conductivity $\kappa_{xy}$, and the thermal Hall angle $\tan \theta_{\rm H}$ 
normalized by the value at the transition temperature 
without impurities and external fields $T_{\rm c}$, 
for the different scattering phase shifts $\sin^2 \delta_0 = 0, 0.2, \dots, 1$ 
at the normal relaxation time $\tau_0 = 10 \hbar / \Delta_0(0)$
as a function of temperature. 
We reproduce the longitudinal thermal conductivity $\kappa_{xx}$ 
proposed by Graf {\it et al}. \cite{Graf}, 
and show that the behavior of the normalized thermal Hall conductivity 
is similar to that of the normalized longitudinal thermal conductivity. 
We also observe that the enhancement of the normalized thermal Hall angle near zero temperature 
decreases as $\sin^2 \delta_0$ increases from $\sin^2 \delta_0 = 0$, 
and does not occur at $\sin^2 \delta_0 = 1$. 
Figures \ref{fig11} and \ref{fig12} plot the thermal Hall angle 
normalized by the value at the transition temperature 
without impurities and external fields $T_{\rm c}$, 
for the different normal relaxation times 
$\tau_0 = 1.5 \hbar / \Delta_0(0), 2 \hbar / \Delta_0(0), \dots, 100 \hbar / \Delta_0(0)$ 
in the Born and unitarity limit as $\sin^2 \delta_0 = 0$ and $1$, respectively, 
as a function of temperature. 
It is shown that the normalized thermal Hall angle is greatly enhanced 
in the Born limit as $\sin^2 \delta_0 = 0$ near zero temperature 
but is suppressed in the unitarity limit as $\sin^2 \delta_0 = 1$ near zero temperature 
even at the large normal relaxation time $\tau_0 = 100 \hbar / \Delta_0(0)$. 
We can explain the normal relaxation time and temperature dependence 
of the normalized thermal Hall angle in the Born and unitarity limit as follows, 
using the corresponding quasiparticle DOS in Figs. \ref{fig13} and \ref{fig14}.

We first consider the Born-type impurity as $\sin^2 \delta_0 = 0$.  
Then the quasiparticle relaxation time $\tau_{\rm  QP}$ can be written from Eq. (\ref{kappa-d}) as 
$\tau_{\rm  QP} = \tau_0 / \langle {\rm Re}g_0^{\rm R} \rangle_{\rm F}$
near zero temperature, 
since quasiparticles around the gap nodes at $\phi = 0$ become dominant. 
Furthermore, in the $d$-wave pairing case, 
since the quasiparticle DOS has a finite small value near $\varepsilon = 0$, 
the impurity scattering for quasiparticles near $\varepsilon = 0$ is very small. 
We can also explain that  
the quasiparticle impurity scattering is very small near zero temperature, 
since quasiparticles around the gap nodes 
are restricted to the momentum in a specific orientation. 
In the Born limit, the normalized thermal Hall angle near zero temperature can be roughly written as
\begin{align}
\frac{\tan \theta_{H}(T)}{\tan \theta_{H}(T_{\rm c})} \sim 
\int d \varepsilon \frac{1}{\langle {\rm Re}g_0^{\rm R} \rangle_{\rm F}} {\rm sech}^2 \frac{\varepsilon}{2 k_{\rm B} T}, 
\end{align}
where the $\varepsilon$ integration signifies an integration over $- \infty \le \varepsilon \le \infty$, 
except $\langle {\rm Re}g_0^{\rm R} \rangle_{\rm F} = 0$. 
Thus, the normalized thermal Hall angle is more enhanced in the Born limit 
at the large normal relaxation time near zero temperature, 
since the quasiparticle DOS near $\varepsilon = 0$ becomes small as the normal relaxation time increases,  
as seen in Fig. \ref{fig13}. 
On the other hand, when taking the unitarity limit as $\sin^2 \delta_0 = 1$, 
the quasiparticle relaxation time $\tau_{\rm  QP}$ near zero temperature can be given by 
$\tau_{\rm  QP} = \tau_0 \langle {\rm Re}g_0^{\rm R} \rangle_{\rm F}$, 
and the restriction on the direction of the quasiparticle momentum is relaxed due to the strong scattering effect. 
In the unitarity limit, the normalized thermal Hall angle near zero temperature can be also roughly written as
\begin{align}
\frac{\tan \theta_{H}(T)}{\tan \theta_{H}(T_{\rm c})} \sim 
\int d \varepsilon \langle {\rm Re}g_0^{\rm R} \rangle_{\rm F} {\rm sech}^2 \frac{\varepsilon}{2 k_{\rm B} T}. 
\end{align}
Thus, the normalized thermal Hall angle in the the unitarity limit is more suppressed 
at the large normal-relaxation-time near zero temperature
inversely to the Born-type impurity, since the quasiparticle DOS near $\varepsilon = 0$ is smaller 
as $\tau_0$ increases, as seen Fig. \ref{fig14}. 
However, 
since the thermal conductivity in $d$-wave superconductors with scattering close to the unitarity limit 
is also almost zero at low temperature, 
it may be difficult to observe the suppression of the Hall angle experimentally.

In our results, the normalized thermal Hall angle near zero temperature
was up to about $100$ times larger than that at the transition temperature. 
From the figures in Refs. \onlinecite{Zhang01} and \onlinecite{Kasahara} 
estimating roughly the ratio of the thermal Hall angle 
at zero temperature to that at the transition temperature,  
that in Ref. \onlinecite{Zhang01} is about 100, 
and that in Ref. \onlinecite{Kasahara} is about 15. 
Here we note that these thermal-Hall-angle ratios were estimated in the limit of zero magnetic field.
The thermal-Hall-angle ratio in Ref. \onlinecite{Zhang01} 
is approximately consistent with our result for the large normal relaxation time in the Born limit, 
and that in Ref. \onlinecite{Kasahara} 
is almost the same as the result for 
$\tau_0 = 100 \hbar / \Delta_0(0)$ and $\sin^2 \delta_0 = 0.8$, 
or $\tau_0 = 10 \hbar / \Delta_0(0)$ and $\sin^2 \delta_0 = 0.6$, 
or \dots, 
or $\tau_0 = 2 \hbar / \Delta_0(0)$ and $\sin^2 \delta_0 = 0$. 
Thus, we find from calculations of the thermal Hall angle that 
YBCO in Ref. \onlinecite{Zhang01} is very clean, 
and materials in Ref. \onlinecite{Kasahara} is relatively dirty 
within this our present theory.

\section{Concluding remarks \label{sec:V}}
We derived the thermal conductivity in extreme type-II superconductors 
with an isolated pinned vortex
based on the augmented quasiclassical equations of superconductivity 
with the Lorentz force. 
Using it, we calculated the thermal conductivity and the thermal Hall angle 
in $d$-wave superconductors, 
and addressed the enhancement of $\lim_{H \to 0} \tan \theta_{\rm H} / H$ 
at low temperatures in YBCO and CeCoIn$_5$ measured 
by Zhang {\it et al.} \cite{Zhang01} and by Kasahara {\it et al.} \cite{Kasahara}. 
They estimated $\lim_{H \to 0} \tan \theta_{\rm H} / H$ 
using the initial slope in the $H$ dependence of $\kappa_{xy}$ measured experimentally.
We also should point out that since the superfluid velocity ${\bm v}_{\rm s}$ is proportional to $1/\rho$, 
the Doppler shift effect has the potential to contribute to the spatial average of the thermal conductivity 
and the thermal Hall angle. 
Thus, we calculated the magnetic-field dependence of $\tan \theta_{\rm H}$ and $\kappa_{xy}$ 
due to the Doppler shift effect, 
and confirmed that the Doppler shift effect can be neglected in the limit $H \to 0$.

The thermal Hall conductivity due to the vortex scattering at weak magnetic fields 
proposed by Durst {\it et al.} in Ref. \onlinecite{Durst} 
is given by $\kappa_{xy}^{\rm vortex} = C_0 T \sqrt{H}$,  
where $C_0$ is a constant introduced in Ref. \onlinecite{Durst}, 
and it has been used successfully 
to explain the experimental results on the magnetic field dependence of the thermal conductivity in 
YBa$_2$Cu$_3$O$_{6.99}$ in the weak magnetic-field region. 
However, they did not address the enhancement of $\lim_{H \to 0} \tan \theta_{\rm H} / H$ 
at low temperatures 
even in Refs. \onlinecite{Ganeshan} and \onlinecite{Kulkarni}, which are their more recent papers, 
and we cannot explain this enhancement using their theories. 
Unlike our theory, 
since their theory takes into account only the contribution of quasiparticles around the gap nodes, 
it is impossible to investigate 
the wide temperature dependence from the transition temperature to zero temperature. 
Moreover, $\kappa_{xy}^{\rm vortex}$ is proportional to $\sqrt{H}$, 
and $\kappa_{xy}^{\rm vortex}/H$ diverges for $H \to 0$, 
i.e., the quasiparticle relaxation time due to the vortex scattering diverges for $H \to 0$
regardless of temperature, 
since the longitudinal thermal conductivity becomes $\kappa_{xx} \sim T$ for $H \to 0$. 
On the other hand, the thermal Hall conductivity due to the Lorentz force 
$\kappa_{xy}^{\rm Lorentz}$ is proportional to $H$ [see Eq. (\ref{kappa})], 
and $\kappa_{xy}^{\rm Lorentz}/H$ has a finite value for $H \to 0$. 
Therefore, $\lim_{H \to 0} \tan \theta_{\rm H} / H$ and $\lim_{H \to 0} \kappa_{xy} / H$ 
are dominated by the Lorentz force and the impurities. 
We observed that 
the thermal Hall angle is greatly enhanced in $d$-wave superconductors 
without impurities of the resonant scattering 
at low temperatures and very weak magnetic field. 
This great enhancement of the thermal Hall angle has been observed 
experimentally in YBCO \cite{Zhang01} and CeCoIn$_5$ \cite{Kasahara}, 
and may also be observed experimentally in other nodal superconductors 
with large magnetic-penetration-depth.

On the other hand, 
our absolute values of the thermal conductivity are different from the experimental values.
To discuss the experimental values more quantitatively, 
we may need to consider inelastic scattering
as considered in the calculation of the longitudinal component, 
such as scattering by the antiferromagnetic spin fluctuations 
within the RPA \cite{Hirschfeld96}
or the fluctuation exchange approximation \cite{Hara} in YBCO. 
The relaxation times for inelastic scattering in the superconducting phase of UPt$_3$ and CeCoIn$_5$ 
are also given in Refs. \onlinecite{Fledderjohann} and \onlinecite{Graf96}, and \onlinecite{Truncik}. 
Here we emphasize that the great enhancement of the electrical and longitudinal thermal conductivity 
in the superconducting state of YBCO is dominated 
by the contribution from inelastic scattering \cite{Hirschfeld96}, 
but thermal Hall angle is greatly enhanced in nodal superconductors even without inelastic scattering.

We also discuss the contribution from quasiparticles in the vortex core. 
For clean superconductors in the vortex lattice state under larger magnetic field, 
we should consider the contribution of quasiparticles in the vortex core. 
The thermal Hall angle in the vortex lattice state may decrease
due to the increase in quasiparticle scattering in the vortex core
as the external magnetic field increases \cite{Kasahara}, 
but we also need to calculate it microscopically and quantitatively
to clarify the suppression of the thermal Hall angle due to external magnetic field.
We may confirm this 
by solving the augmented quasiclassical equations directly in the vortex lattice system \cite{Adachi,Ichioka99}
(see Ref. \onlinecite{Ueki18} for the pair-potential-gradient force and 
the pressure difference arising from the slope in the DOS). 
This method can include not only the contribution of quasiparticles outside the core 
but also that of the Andreev-reflected quasiparticles \cite{Dahm,Nagai}
in the thermal conductivity. 
We may also calculate the thermal Hall conductivity in high magnetic fields more easily, 
using the Brandt--Pesch--Tewordt approximation 
for the augmented quasiclassical equations \cite{Houghton}.
Moreover, we may also need to confirm the contributions from skew scattering due to a vortex \cite{Durst,Kulkarni} 
and the Berry phase acquired by a quasiparticle on traversing around a vortex 
\cite{Vafek01,Cvetkovic,Murray,Vafek15,Ganeshan}
in the calculations under larger magnetic field.

Our approach for the study of thermal Hall conductivity in type-II superconductors 
can be applied to the microscopic study of quasiparticle transport in a variety of 
superconductors under magnetic field.

\begin{acknowledgments}
H. U. thanks K. Izawa and J. Goryo 
for extensive discussions on the thermal conductivity in superconductors, 
Y. Masaki and W. Kohno for sound advice about the impurity and Volovik effects, 
T. Matsushita and A. Kirikoshi  for helpful discussions on the formulation in this work, 
and E. S. Joshua for many discussions and comments on our present paper. 
This work was partially supported by JSPS KAKENHI Grant No. 15H05885 (J-Physics). 
The computation in this work was partially carried out using the facilities of the Supercomputer Center, 
the Institute for Solid State Physics, the University of Tokyo.
\end{acknowledgments}

\end{document}